\documentclass[fleqn,usenatbib]{mnras}

\usepackage{newtxtext,newtxmath}

\usepackage[T1]{fontenc}

\DeclareRobustCommand{\VAN}[3]{#2}
\let\VANthebibliography\thebibliography
\def\thebibliography{\DeclareRobustCommand{\VAN}[3]{##3}\VANthebibliography}

\usepackage{graphicx}	
\usepackage{amsmath}	

\title[Black Hole Spin Mass-Energy Characteristics]
{Robust Supermassive Black Hole Spin Mass-Energy Characteristics: A New Method and Results}

\author[R. A. Daly]{Ruth A. Daly\thanks{E-mail: rdaly@psu.edu}\\
Penn State University, Berks Campus, Reading, PA 19608, USA\\}

\date{Accepted XXX. Received YYY; in original form ZZZ}

\pubyear{2022}

\begin{document}
\label{firstpage}
\pagerange{\pageref{firstpage}--\pageref{lastpage}}
\maketitle

\begin{abstract}
The rotational properties of astrophysical black holes are fundamental quantities that characterization the black holes. A new method to empirically determine the spin mass-energy characteristics of astrophysical black holes is presented and applied here. Results are obtained for a sample of 100 supermassive black holes with collimated dual outflows and redshifts between about zero and two. An analysis indicates that about two-thirds of the black holes are maximally spinning, while one-third have a broad distribution of spin values; it is shown that the same distributions describe the quantity $\rm{(M_{rot}/M_{irr})}$. The new method is applied to obtain the black hole spin mass-energy, $\rm{M_{spin}}$, available for extraction relative to: the maximum possible value, the irreducible black hole mass, and the total black hole mass, $\rm{M_{dyn}}$. The total energy removed from the black hole system and deposited into the circumgalactic medium via dual outflows over the entire outflow lifetime of the source, $\rm{E_T}$, is studied relative to $\rm{M_{dyn}}$ and relative to the spin energy available per black hole, $\rm{E_{spin}/(M_{\odot}c^2)}$. The mean value of $\rm{Log(E_T/M_{dyn})}$ is about $(-2.47\pm 0.27)$. Several explanations of this and related results are discussed. For example, the energy input to the ambient gas from the outflow could turn off the accretion, or the impact of the black hole mass loss on the system could destabilize and terminate the outflow. The small values and restricted range of values of $\rm{Log(E_T/M_{dyn})}$  and $\rm{Log(E_T/E_{spin})}$   could suggest that these are fundamental properties of the primary process responsible for producing the dual collimated outflows.
\end{abstract}




\section{Introduction} \label{sec:intro}
Black holes are ubiquitous in the universe. Supermassive black holes reside at the 
centers of galaxies and stellar-mass black holes populate galaxies. The two primary 
characteristics that describe an astrophysical black hole are the mass and spin 
of the hole (assuming the black hole has negligible charge). Often in astrophysical contexts, the mass of a black hole 
is empirically determined by the dynamics 
and properties of matter and light in the vicinity of the black hole. 
The black hole mass that will be measured is the total 
black hole mass, $\rm{M}$, which has contributions from the 
irreducible mass, $\rm{M_{irr}}$, and the mass-energy associated with the 
spin angular momentum of the black hole $J$: $\rm{M = (M_{irr}^2 + (Jc/(2GM_{irr})^2)^{1/2}}$, where G is the gravitational constant 
and $c$ is the speed of light (e.g. Christodoulou 1970; Bardeen, Press, \& Teukolsky 1972; 
Misner, Thorne, \& Wheeler 1973; Rees 1984; Blandford 1990); this equation may be rewritten in the form 
of eqs. (3) and (9). 
The mass-energy that can be extracted from the spinning black hole, referred to here as the "spin mass-energy," is $\rm{M_{spin} = M - M_{irr}}$, as described in detail by Thorne et al. (1986) (see their eq. 3.88 and related discussion). Thus, there are two different quantities that have been referred to as spin or rotational mass-energy of the black hole in the literature 
(e.g. Rees 1984; Thorne et al. 1986; Gerosa, Fabbri, \& Sperhake 2022). 
For clarity, throughout this paper, the quantity 
$\rm{M_{rot} \equiv Jc/(2GM_{irr})}$ is 
referred to as the "rotational mass." The quantity 
$\rm{M_{spin} \equiv M - M_{irr} \equiv E_{spin} c^{-2}} $ is referred to as the 
"spin mass-energy" of the black hole and indicates the 
mass-energy that is available to be extracted from the black hole (see, for example,  
Blandford \& Znajek 1977; Rees 1984; Thorne et al. 1986, 
and Blandford 1990 for detailed discussions). 
The mass, $\rm{M}$, is also referred to as the dynamical mass, $\rm{M_{dyn}}$, since it  
is the total black hole mass that will be inferred by dynamical and other astronomical studies.

The irreducible mass of an isolated black hole can not be reduced or decreased, 
but mass-energy associated with black hole spin can be 
extracted, thereby decreasing the total mass of the hole 
(Penrose 1969; Penrose \& Floyd 1971; Blandford \& Znajek 1977). 
Collimated outflows from 
supermassive black holes associated with 
active galactic nuclei (AGN) and stellar-mass black holes associated with 
X-ray binaries are likely to be powered, at least in part, by black hole spin 
(e.g. Blandford \& Znajek 1977; MacDonald \& Thorne 1982; Phinney 1983; 
Begelman, Blandford, \& Rees 1984; Blandford 1990; Daly 1994, 1995; Moderski, Sikora, 
\& Lasota 1998; Meier 1999; Koide et al. 2000; Wan et al. 2000; 
Punsly 2001; Daly \& Guerra 2002; 
De Villiers, Hawley, \& Krolik 2003; Gammie et al 2004; 
Komissarov \& McKinney 2007; Beckwith, Hawley, \& Krolik 2008; 
King, Pringle, \& Hofmann 2008; Miller et al. 2009; 
O'Dea et al. 2009; Daly 2009a,b; Tchekhovskoy et al. 2010; Daly 2011; 
Gnedin et al. 2012; King et al. 2013; Ghisellini et al. 2014; Yuan 
\& Narayan 2014; Daly \& Sprinkle 2014; Daly 2016; 
Gardner \& Done 2018; Krause et al. 2019; 
Reynolds 2019; Daly 2019). In this case, the spin energy extracted during the outflow will 
cause the black hole mass to decrease. A source that undergoes multiple outflow 
events could significantly drain the spin energy of the hole and thereby 
decrease the black hole mass. 
The amount of spin energy extracted during outflow events have been estimated 
for radio sources with large-scale outflows such as 
FRI sources in galaxy-cluster environments 
(McNamara et al. 2009; Daly 2009a,b, 2011), FRII sources 
(Daly 2009a,b; 2011), and several types of AGN and stellar-mass 
black holes (Daly 2020). FRI sources are extended radio sources that are 
"edge-darkened" while FRII sources, also known as classical doubles, 
are "edge-brightened" (Fanaroff \& Riley 1974). 
In addition, the fraction of the spin energy extracted
per outflow event has been estimated for FRII sources (Daly 2011), and 
is roughly a few to several percent. 
Thus, in models in which collimated outflows from the vicinity of
a black hole are powered by black hole spin, the spin and spin-energy of the hole 
are expected to decrease as a result of the outflow. 

The fact that the mass-energy associated with black hole spin may 
be extracted, modified, or reduced, 
and thus that the total or dynamical mass of a black hole can be 
reduced may introduce 
dispersion in relationships between black hole mass and properties of the host galaxy 
(e.g. Kormendy 
\& Richstone 1995; Ferrarese \& Ford 2005; Kormendy \& Ho 2013; Shankar 2013; 
Sesana et al. 2014; Zubovas \& King 2019; King \& Nealon 2019; King \& Pounds 2015). 
If black hole spin evolves with 
redshift, this is likely to  cause an evolution in these relationships and 
their dispersion. Additionally, black hole spin is expected to evolve with redshift 
as a result of the merger and accretion history of the black hole 
(e.g. Hughes \& Blandford 2003; Gammie et al. 2004; 
Volonteri et al. 2005, 2007; King \& Pringle 2006, 2007; King et al. 2008;
Berti \& Volonteri 2008; Ghisellini et al. 2013). 
Thus, the study of black hole spin evolution provides insight into the 
merger and accretion history of supermassive black holes. 
Black hole spin may depend upon galaxy type or environment (e.g. Sesana et al. 2014; 
Antonini et al. 2015; King \& Pounds 2015; Barausse et al.  2017; 
King \& Nealon 2019), which may lead to environmental changes in the relationship between 
black hole mass and galaxy properties, 
or a change in the dispersion of relationships (e.g. Zubovas \& King 2012). 
The dispersion introduced 
may be complex and will depend upon the initial spin and irreducible 
mass of the black hole, the processes responsible for spinning up the hole 
such as accretion 
or mergers, processes which tap or reduce the spin of the hole, 
and the complex interaction of feedback, accretion, outflows, and other processes 
associated with the black hole, which are likely to play a role in determining the spin 
and thus spin mass-energy and dynamical mass of the hole 
(e.g. Belsole et al. 2007; Worrall 2009; Voit et al. 2015; 
Hardcastle \& Croston 2020). In addition, 
it is likely that some sources undergo multiple outflow events (e.g. Hardcastle et al. 
2019; Bruni et al. 2019, 2020; Shabala et al. 2020), 
so that even if a small amount of the spin energy is extracted 
per outflow event, over time a substantial amount of spin energy can 
be extracted due to multiple outflow events. 

The distinction between dynamical mass, spin mass-energy, and irreducible mass of a  
black hole is also important when comparing empirically determined quantities with 
theoretically predicted quantities, such as those indicated by numerical simulations. 
Numerical simulations predict the expected black hole spin and mass evolution in 
the context of different black hole merger and accretion histories (e.g. 
King et al. 2008; Volonteri et al. 2013; 
Dubois, Volonteri, \& Silk 2014; Sesana et al. 2014; Kulier et al. 2015). 
A comparison of simulation results with empirically determined results  
provides an important diagnostic of the merger and accretion histories of 
black holes located at the centers of galaxies. 

The number of available black hole spin values, and therefore black hole spin
energies, has recently increased substantially. 
The development of the "outflow method" of empirically determining black hole 
spin and accretion disk properties developed and described by Daly (2016) and 
Daly (2019)  
(hereafter D16 and D19) and 
Daly et al. (2018), allow the empirical determination of the black hole spin 
function, spin, and, accretion disk properties such as the 
mass accretion rate and disk magnetic field strength for over 750 sources.  
D19 showed that the fundamental equation that describes an outflow 
powered at least in part by black hole spin,
$\rm{L_j \propto B_p^2 \rm{M_{dyn}}^2 F^2}$ (e.g. Blandford \& Znajek 1977; 
Meier 1999; Tchekhovskoy et al. 2010; Yuan \& Narayan 2014) is 
separable and may be written as 
\begin{equation}
\rm{(L_j/L_{Edd}) = g_j ~(B/B_{Edd})^2 ~F^2}
\end{equation}
(see eq. 6 from D19); here $\rm{B_p}$ is the poloidal component of the accretion disk magnetic field, 
$\rm{B}$ is the magnitude of disk magnetic field, $\rm{B_{Edd}}$ is the Eddington magnetic 
field strength (e.g. Rees 1984; Blandford 1990; Dermer et al. 2008; D19), 
$\rm{B_{Edd}} \approx 6 \times 10^4 (\rm{M_{dyn}}/10^8 M_{\odot})^{-1/2}$ G, $\rm{F^2} \equiv \rm{f(j)/f_{max}}$ 
is the normalized spin function (discussed in more detail in section 2), $\rm{f_{max}}$ 
is the maximum value of  the spin function $\rm{f(j)}$, and 
$\rm{g_j}$ is the normalization factor for the beam power $\rm{L_j}$ in units of the 
Eddington Luminosity, $\rm{L_{Edd}}$, $\rm{(L_j/L_{Edd})(max) = g_j}$. 
Note that the ratio $\rm{(B_p/B)^2}$ is absorbed into the 
normalization factor $\rm{g_j}$, thus $\rm{g_j}$ may depend upon AGN type, as 
discussed in sections 3.1 and 4 of D19. 
(Also note that even though the maximum value of the spin function $\rm{f(j=1) = f_{max}} = 1$, 
the normalization term $\rm{f_{max}}$ is included in eqs. (1) and (7) for completeness 
since in some numerical simulations 
$\rm{f(j)}$ is described by modified representations  
(e.g. Tchekhovskoy et al. 2010). Here, since $\rm{f_{max}} = 1$, the terms "spin function" and 
"normalized spin function" are used interchangeably.) 

The spin functions obtained by D19 were converted to 
dimensionless black hole spin angular momentum values, 
and compared with values obtained with independent methods such as those discussed by Azadi et al. (2020) and Reynolds (2019);  
see also, for example, Gnedin et al. (2012), Patrick et al. (2012),   
King et al. (2013), Walton et al. (2013), Wang et al. (2014), 
Garc\'ia et al. (2015), Mikhailov et al. 
(2015, 2019), Vasudevan et al. (2016), Piotrovich et al. (2017, 2020), and 
Mikhailov \& Gnedin (2018). A comparison of 
black hole spin parameters obtained independently with the outflow method and the continuum fitting method was possible for 15 of the sources 
studied with both methods (Azadi et al. 2020), and consistent spin 
parameters were obtained with the two methods. And, a comparison was 
possible and very good agreement was found for six AGN and one stellar mass
black hole studied with both the outflow method and 
the X-ray reflection method (e.g. Fabian et al. 1989; 
Iwasawa et al. 1997; Miller et al. 2002; Reynolds 2019), which included 
all of the sources for which a comparison is currently possible. 
Thus, all sources for which independent spin angular momentum 
values could be compared indicate good agreement between 
independently determined values. 
The high spin values obtained are also consistent
with expectations based on AGN luminosities 
(e.g. Sun \& Malkan 1989; Davis \& Laor 2011; Wu et al. 2013; Trakhtenbrot 2014; 
Brandt \& Alexander 2015; Trakhtenbrot, Volonteri, \& Natarajan 2017). 
So, the expectation is that some significant fraction of black holes are likely 
to have high spin; there may also be a population of black holes with lower 
spin, and, of course, black hole spin is likely to be an evolving quantity. 

Here, a new method to study the spin mass-energy characteristics of black holes is 
presented and applied to a sample of 100 supermassive black holes 
with empirically determined black hole spin functions. 
This is important because the spin mass-energy of a black hole 
can be extracted, thereby reducing the total black hole mass, and 
energy channelled away from the hole can significantly 
impact the near and far field environments of the black hole. 
The traditional and new methods of empirically determining 
the spin mass-energy characteristics of black holes 
are described in sections 2.1 and 2.2, respectively.  

Empirically determined spin functions, $F^2$, are used to 
obtain the spin mass-energy characteristics 
of the sample of 100 supermassive black holes, bypassing the use
of dimensionless spin angular momenta, $j$. 
The properties of the spin functions (see eq. 7) 
are described and analyzed in section 2.3, and it is found that  
the sources are well described by  
a population of maximally spinning black holes plus a 
population of holes with a broad distribution of spin values. 
In section 3, the empirically determined spin functions are applied to 
obtain for each black hole: 
the spin mass-energy relative to the maximum possible value;  
the spin mass-energy relative to the 
irreducible and dynamical black hole mass;  
the spin mass-energy in units of solar masses; the ratio of the total outflow energy to the black hole 
spin energy; the ratio of the total outflow energy to the dynamical  
black hole mass; and the ratios of the rotational mass relative to the 
irreducible and dynamical black hole masses.  
The results are discussed and summarized in sections 4 
and 5. 

All quantities are obtained in a spatially flat cosmological model
with two components, a mean mass density relative to the critical
value at the current epoch of $\Omega_m = 0.3$ and  
a similarly normalized cosmological constant of
$\Omega_{\Lambda} = 0.7$. A value for 
Hubble’s constant of $H_0 = 70 ~\rm{km~ s}^{-1} \rm{Mpc}^{-1}$ is  
assumed throughout. 

\section{Method}
The traditional method of obtaining black hole spin mass-energy 
characteristics and some of the difficulties that arise in the application of this method to empirically determine spin mass-energy properties of astrophysical black holes 
are described in section 2.1. The new method avoids these difficulties by characterizing the spin mass-energy characteristics in terms 
of the spin function; the new method that will be 
applied here is presented in section 2.2. The properties of the 
empirically determined spin functions that will be applied to 
obtain and study the spin mass-energy characteristics of 100 supermassive 
black holes are discussed in section 2.3.

\subsection{The Traditional Method}
As described in section 1, the rotational energy of a black hole contributes to the total dynamical black hole mass, $\rm{M_{dyn}}$, 
which is the mass that will be measured by a distant observer, 
and $\rm{M_{dyn}^2 = M_{irr}^2 + M_{rot}^2}$ (see eq. 3), 
where $\rm{M_{rot} \equiv (Jc/(2GM_{irr}))}$. 
The spin energy $\rm{E_{spin}}$ that may be extracted is $\rm{E_{spin} = M_{spin} c^2}$, 
where $\rm{M_{spin} = M_{dyn} - M_{irr}}$, 
and is referred to here as the black hole spin energy or spin mass-energy.

Relationships between the dimensionless black hole spin angular momentum, $j$, the total
black hole mass, $\rm{M}$ (also referred to as $\rm{M_{dyn}}$),  
and the mass-energy that can be extracted from the black hole,  
$\rm{M_{spin}}$, are   
discussed, for example, by  Misner, Thorne, \& Wheeler (1973), Rees (1984), 
Blandford (1990), and 
Thorne et al. (1986). The dimensionless black hole spin angular momentum $j$ is defined in the usual way in terms of the spin angular momentum $J$ and the total black hole mass 
$M$, 
$j \equiv Jc/(G M^2)$; in other work, $j$ is sometimes represented with the 
symbol $a_*$ or $a/M$. As described in section 1, the work of Thorne et al. 
(1986) (see also Rees 1984; Blandford 1990), indicates the following set of equations:  
\begin{equation}
 \rm{ M \equiv M_{dyn}} = \rm{M_{irr}} + \rm{E_{spin}} c^{-2} = \rm{M_{irr}} + \rm{M_{spin}}  
\end{equation}
and 
\begin{equation}
\rm{M_{irr}} = \rm{M_{dyn}} \left({{1 + (1-j^2)^{1/2}} \over 2}\right)^{1/2}~,
\end{equation}
where eq. (3) follows from $\rm{M_{dyn}^2 = M_{irr}^2 + (Jc/(2GM_{irr}))^2}$, 
discussed in section 1. Eqs. (2) and (3) indicate that 
\begin{equation}
{\rm{M_{spin}} \over \rm{M_{dyn}}} = 1 - \left({\rm{M_{irr}} \over \rm{M_{dyn}}}\right)
\end{equation}
and 
\begin{equation}
{\rm{M_{spin}} \over \rm{M_{irr}}} = \left({\rm{M_{dyn}} \over \rm{M_{irr}}}\right) -1~.
\end{equation}
Eqs. (3) \& (5) indicate that
\begin{equation}
{\rm{E_{spin}} \over E_{spin,max}} = \left({\sqrt{2}(1+ \sqrt{1-j^2})^{-0.5} - 1]
\over \sqrt{2} -1} \right) \approx 2.41 \left(\rm{{M_{spin}} \over {M_{irr}}}\right), 
\end{equation}
where $(\rm{E_{spin}/E_{spin,max}})$ is obtained by dividing 
eq. (5) as a function of j by eq. (5) with $j=1$, since 
$(\rm{E_{spin,max}/M_{irr}})$ is obtained with eqs. (3) and (5) assuming a value of $j=1$.

\begin{figure}
    \centering
    \includegraphics[width=\columnwidth]{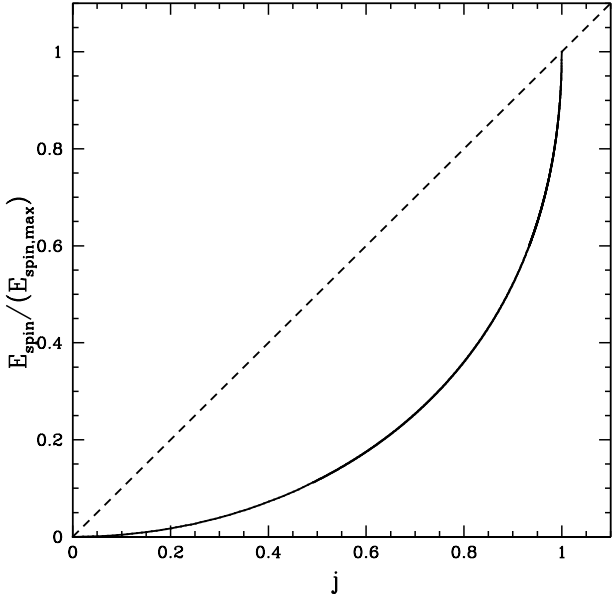}
\caption{The solid line shows the black hole spin energy available for extraction, $\rm{E_{spin}}$ in
units of the maximum possible value of this energy, $\rm{E_{spin,max}}$,  
versus the 
dimensionless black hole spin angular momentum $j$ (defined in section 2.1). The dotted line provides a comparison to a linear relationship.}
		  \label{fig:FEspinoverEspinmax}
    \end{figure} 

\begin{figure}
    \centering
    \includegraphics[width=\columnwidth]{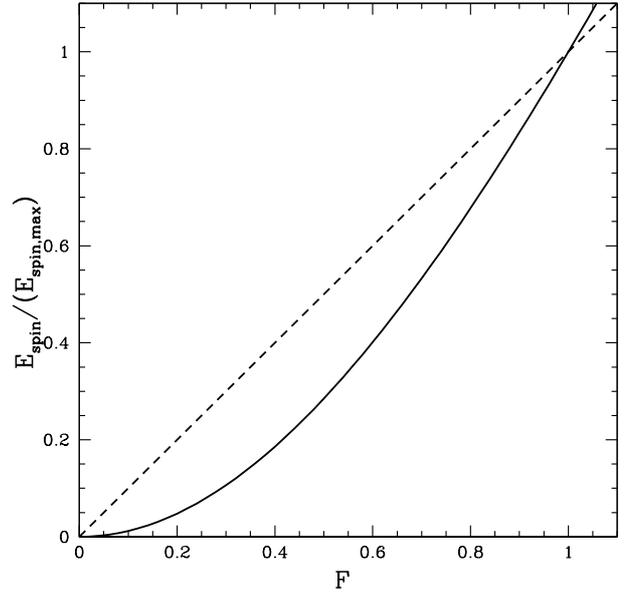}
\caption{The solid line shows  
the available black hole spin energy $\rm{E_{spin}}$ 
normalized to the maximum possible value of the spin energy, $\rm{E_{spin,max}}.$ as a function of $\rm{F}$, the square root of the black hole spin function (defined by eq. 7). The  dotted line provides a comparison to a linear relationship.}
		  \label{fig:Fig2}
    \end{figure} 

There are several factors that indicate it is preferable to rewrite equations (3 - 5) in terms of the spin function $\rm{F^2 \equiv f(j)/f_{max}}$. 
In the application of the outflow method (D16, D19), 
the quantity that is determined empirically is $\rm{F}$, so it is preferable to be able to obtain the quantities on the left hand sides of equations (3-6) directly 
in terms of $\rm{F}$, where 
\begin{equation}
F ~\equiv~ \sqrt{f(j) \over f_{max}}~ = ~{j \over (1+\sqrt{1-j^2})}~.
\end{equation}
To further complicate the use of $j$ to empirically characterize the spin properties of a black hole, the quantity 
$\sqrt{(1-j^2)}$ indicates that values of $j$ that are greater than one cannot be accommodated, though the uncertainties associated with empirically determined black hole spin characteristics should allow for this possibility (e.g. Daly 2020).
In addition, the relationship between the normalized spin energy of the black hole and the dimensionless spin angular momentum $j$ is highly non-linear as indicated by 
eq. (6) and illustrated by Fig. 1. 
The spin energy $E_{1/2}$
is about half of the maximum possible spin energy for $j$ of about 
0.9; the spin energy $E_{1/4}$ is about one quarter of the maximum value 
when $j$ is about 0.7; 
and the spin energy $E_{1/10}$ is about one tenth of the maximum value 
when $j$ is about 0.5.
It is clear that the relationship between the normalized spin energy and $j$ is quite 
non-linear, and 
relatively high dimensionless spin angular momentum 
values of 0.5 and 0.7 indicate relatively low spin energy values 
of only 1/10 and 1/4, respectively, of the maximum possible value. 
Another way to 
state this is that relatively low values of spin energy indicate substantial values of dimensionless 
black hole spin angular momentum $j$. Thus, if a black hole has any spin energy at all, it is 
expected to have a value of $j$ substantially different from zero. 

The facts that the relationship between the dimensionless 
spin angular momentum $j$ and the normalized spin energy 
$\rm{(E_{spin}}/E_{spin,max})$ is highly non-linear, that 
empirically determined values of $j$ cannot exceed unity
due to the term $\sqrt{1-j^2}$ even though 
observational uncertainties require this flexibility, and that 
the empirically determined quantity is $\rm{F}$,  
suggest that some other function  
should be used to empirically determine the spin mass-energy 
properties of black holes. And, as noted earlier, 
it is important to determine 
how much of the dynamical mass could be extracted and thereby 
decrease the black hole mass. Spin
energy 
also indicates the potential impact of the "spin energy reservoir" that is stored in spinning black holes on the near and far field environments of black holes. 

\subsection{The New Method}

In contradistinction to these issues, 
the relationship between both the black hole spin function $F$, $F^2$,   
or $\rm{Log(F)}$   
and the 
black hole mass-energy associated with the spin
do not suffer from the limitations described in section 2.1, 
as illustrated in Figs. 2, 3, and 4. Though theoretically $F$ is not expected to exceed unity, empirically determined values of $F$ will exceed unity due to measurement  uncertainties. There are several additional reasons why it is preferable to use the quantity $F$, $F^2$, 
or $\rm{Log(F)}$: (1) there is no mathematical constraint analogous to that for $j$ (described in section 2.1) 
that requires that $F$ must be less than or equal to one; (2) the relationship between $F$ and spin energy is only slightly non-linear as illustrated in Fig. 2;  
(3) the spin energy in units of the maximum spin energy is very well 
approximated by the value of $F^2$ 
for values of $F^2$ between about zero and 1.5 or so  
as illustrated in Fig. 3. Allowing the exponent of $\rm{F}$ to vary, it is found that 
$\rm{Log(\rm{E_{spin}}/E_{spin,max}}) \approx 1.75~ \rm{Log(F)}$, 
as illustrated in Fig. 4.
It should be noted that equations (9-12) should be used to 
obtain uncertainties for $\rm{(M_{irr}}/\rm{M_{dyn})}$, $\rm{(M_{spin}}/\rm{M_{dyn})}$, $\rm{(M_{spin}}/\rm{M_{irr})}$, 
$\rm{(E_{spin}}/\rm{E_{spin,max})}$ and 
related quantities, which are listed below. 

To re-write equations (3-6) in terms of $F$, we 
manipulate the relationship between $F$ and $j$ obtained and discussed by D19 to obtain   
\begin{equation}
 0.5~ [1+ (1-j^2)^{1/2}] =    (F^2 +1)^{-1}.
\end{equation}
This is then substituted into eq. (3) to obtain: 
\begin{equation}
    {\rm{M_{irr}} \over \rm{M_{dyn}}} =  ~(F^2 +1)^{-1/2}~.  
\end{equation}
Combining eqs. (2) and (9) indicates that 
\begin{equation}
{\rm{M_{spin}} \over \rm{M_{dyn}}} = 1 - \left({\rm{M_{irr}} \over \rm{M_{dyn}}}\right) 
= ~[1-(F^2+1)^{-1/2}],  
\end{equation}
and 
\begin{equation}
{\rm{M_{spin}} \over \rm{M_{irr}}} = \left({\rm{M_{dyn}} \over \rm{M_{irr}}}\right) -1~ 
= ~[(F^2+1)^{1/2} -1].
\end{equation}
Eq. (11) indicates that   
\begin{equation}
{\rm{E_{spin}} \over E_{spin,max}}={\rm{M_{spin}} \over M_{spin,max}} = \left({\sqrt{F^2+1} -1 
\over \sqrt{2} -1}\right) \approx 2.41 \left({\rm{M_{spin}} \over \rm{M_{irr}}}\right)
\end{equation}
where $(\rm{E_{spin}/E_{spin,max}})$ is obtained by dividing 
eq. (11) as a function of F by eq. (11) with $F=1$, since 
$(\rm{E_{spin,max}/M_{irr}})$ is obtained with eqs. (9) and (11) assuming a value of $F=1$.

\begin{figure}
    \centering
    \includegraphics[width=\columnwidth]{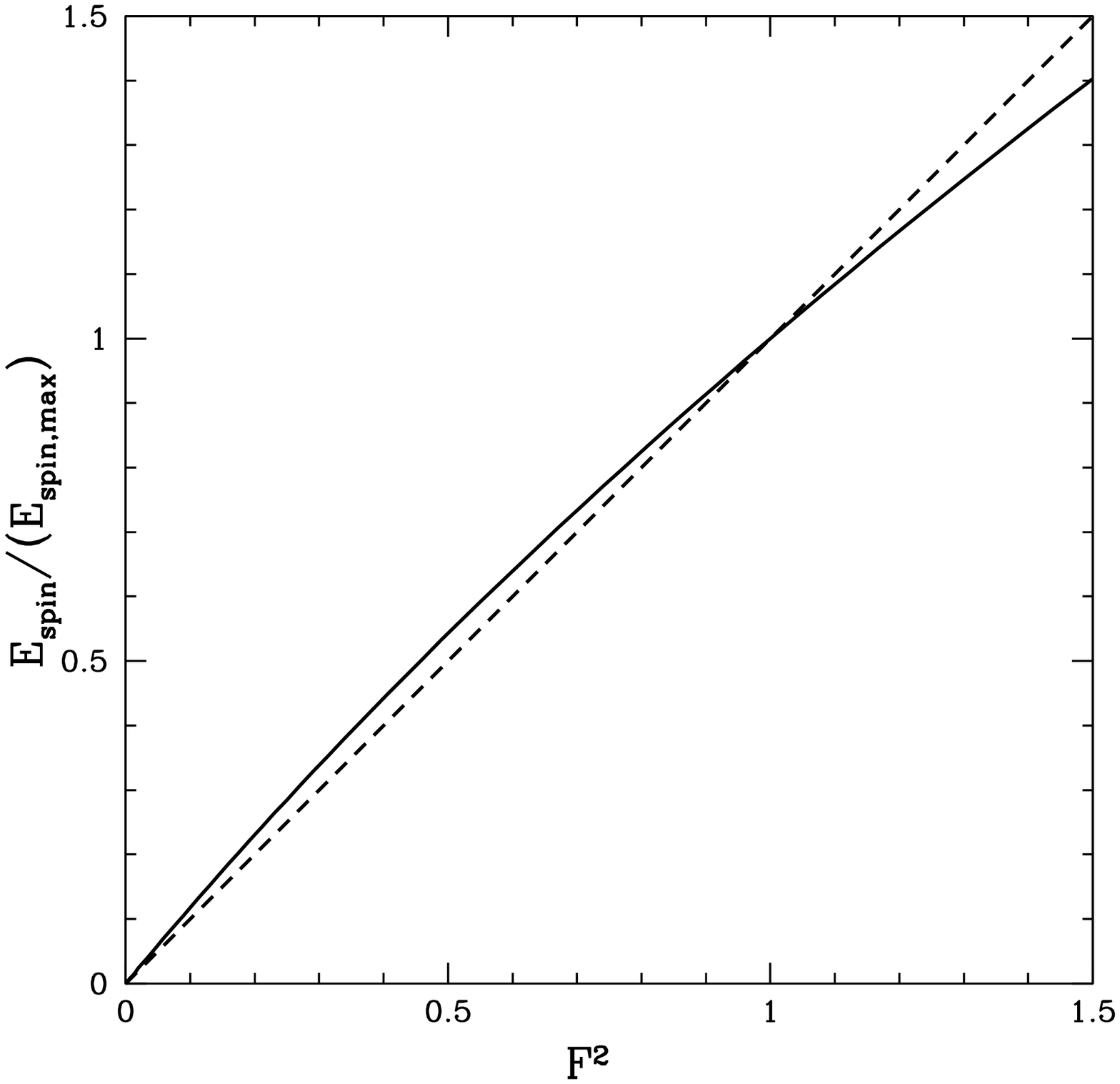}
\caption{The solid line shows the available black hole spin energy $\rm{E_{spin}}$ 
normalized to the maximum possible value of this spin energy, $\rm{E_{spin,max}}$, as a function of the black hole spin function, $\rm{F^2}$ 
(defined by eq. 7). The  dotted line provides a comparison to a linear relationship.}
		  \label{fig:Fig3}
    \end{figure} 
    
    \begin{figure}
    \centering
    \includegraphics[width=\columnwidth]{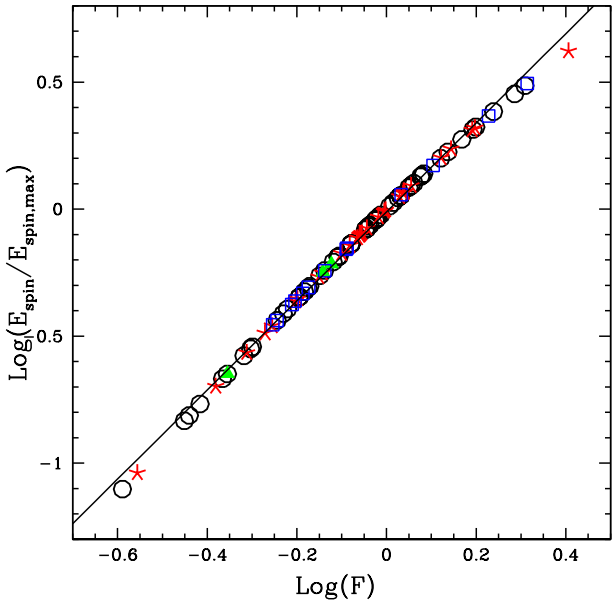}
\caption{$\rm{Log(\rm{E_{spin}}/E_{spin,max})}$ 
as a function of the $\rm{Log(F)}$ is illustrated with input values of $\rm{Log(F)}$ 
selected to match those of the 100 FRII sources that will be considered here, 
but the input values of $\rm{Log(F)}$ could have been selected as in Figs. 2 and 3. 
The unweighted best fit line (solid line) has a slope of $1.75 \pm 0.01$, y-intercept 
of $-0.011 \pm 0.002$, and $\chi^2 = 0.03$. Here the exponent that $\rm{F}$  is raised to is allowed to vary, while in Fig. 3 this exponent is fixed. 
The symbols and colors are as in Fig. 6.}
		  \label{fig:Fig3}
    \end{figure}

Equations (9-12) will be applied to empirically determine the 
ratio of the dynamical black hole mass to the 
irreducible black hole mass, $\rm{M_{dyn}}/\rm{M_{irr}}$, 
the ratio of the spin mass-energy to the dynamical 
mass $\rm{M_{spin}}/\rm{M_{dyn}}$, the ratio of the spin 
mass-energy to the irreducible black hole mass $\rm{M_{spin}}/\rm{M_{irr}}$, 
and the spin energy in terms of the maximum spin energy, 
${\rm{E_{spin}}/E_{spin,max}}$ 
for a sample of 100 FRII sources. Of course, the maximum value of $\rm{M_{spin}}$  
obtained from eqs. (10) and (11) with $F=1$, remain 
$(\rm{M_{spin}}/\rm{M_{dyn}})(max) \simeq 0.29$ and 
$(\rm{M_{spin}}/\rm{M_{irr}})(max) \simeq 0.41$, while the maximum value of 
$(\rm{M_{dyn}}/\rm{M_{irr}})$ obtained with $F=1$ (or $j = 1$) is $\sqrt{2}$. 

Eqs. (9-12) indicate the following uncertainties:  
$\delta(\rm{M_{dyn}}/\rm{M_{irr}}) = F [(F^2+1)^{-0.5}]~ \delta{F}$; 
$\delta(\rm{M_{spin}}/\rm{M_{dyn}}) = F [(F^2+1)^{-1.5}] ~\delta{F}$; 
$\delta(\rm{M_{spin}}/\rm{M_{irr}}) = \delta(\rm{M_{dyn}}/\rm{M_{irr}})$, and 
$\delta(\rm{E_{spin}}/E_{spin,max}) = F (F^2+1)^{-0.5} (\sqrt{2}-1)^{-1} \delta F$.
Of course, $\rm{\delta(Log(x)) = (\delta(x)/x)/ln}(10)$. These 
uncertainties are included in the Tables and shown in plots of 
quantities versus redshift.

The total mass-energy associated with the spin of the black hole can 
be obtained by multiplying $\rm{(M_{spin}}/\rm{M_{dyn})}$ by the empirically determined 
mass of the black hole, $\rm{M}$. This is the first time the 
empirically determined black hole mass is required. Clearly,  
\begin{equation}
{\rm{E_{spin}} \over (M_{\odot} c^2)} = {\rm{M_{spin}} \over M_{\odot}} = M \times \left({\rm{M_{spin}} \over \rm{M_{dyn}}}\right) .
\end{equation}
All empirically 
determined black hole masses are dynamical black hole masses. This will be discussed in more detail in section 4. Empirically determined 
black hole masses and their uncertainties are indicated by the 
Eddington luminosities listed in D16 \& D19. 

The rotational mass, defined and described in section 1, can also be represented in terms 
of the spin function, $\rm{F^2}$. It is easy to show that 
\begin{equation}
\rm{{M_{rot} \over M_{irr}} = F}  
    \end{equation}
    and
    \begin{equation}
\rm{M_{rot} \over M_{dyn}} = {F \over \sqrt{F^2 +1}}. 
    \end{equation}
Each of these quantities can easily be obtained for the 100 sources studied here
from Tables 1 and 2, given that $\rm{Log(M_{rot}/M_{irr}) = Log(F)}$ and 
$\rm{Log(M_{rot}/M_{dyn})} = \rm{[Log(F) - Log(M_{dyn}/M_{irr})]}$, 
as discussed in section 4.1.

\subsection{Properties of the Spin Function}
Spin functions, $F^2$, for the 100 supermassive black holes associated with 
FRII sources obtained by D19 are considered here. The values of 
$\rm{Log(F)}$ and their uncertainties are listed in Tables 1 and 2, and a histogram 
of values is shown in Fig. 5. 
The sources are drawn from the flux limited 3CRR catalogue of radio sources 
(Laing, Riley \& Longair 1983), and thus are subject to well-known selection 
effects such as the loss of lower-luminosity sources as source redshift increases 
(e.g. see Fig. 1 of McLure et al. 2004). To illustrate the impact of this selection 
effect on the histogram of each quantity, the redshift distribution of each quantity 
is provided and discussed; for example, the redshift distribution of  
$\rm{Log(F)}$ is shown in Fig. 6, and will be discussed below. 

\begin{figure}
    \centering
    \includegraphics[width=\columnwidth]{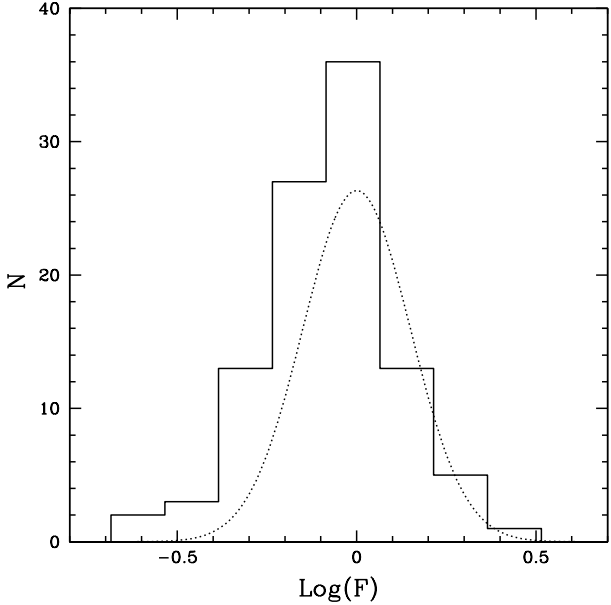}
\caption{The 
histogram of $\rm{Log(F)}$ is shown as the solid line.  The population is well described with a two component model: a population of maximally spinning black holes with 
$\rm{Log(F)} = 0$ and standard deviation $\sigma = 0.15$, 
illustrated by the Gaussian (dotted line), plus a population with 
$-0.6 < \rm{Log(F)} < 0$ with a tilted distribution (see section 2.3). 
Of the sample of 100 black holes studied, about 2/3 are maximally spinning, 
and about 1/3 have a slowly declining distribution of spin functions 
toward lower values of $\rm{Log(F)}$. $\rm{Log(M_{rot}/M_{irr})} = \rm{Log(F)}$ 
(see eq. 14), so this is also the distribution of the 
values of $\rm{Log(M_{rot}/M_{irr})}$. 
For all histograms, 
the bin size is selected to be close to the mean value of the uncertainty of the 
quantity listed in Tables 1 and 2.}
		  \label{fig:Fig14}
    \end{figure} 

The values of $\rm{Log(F)}$ are illustrated with the histogram shown in Fig. 5. The 
bin size in this and all subsequent histograms is selected to be close to the 
mean value of the uncertainty for the quantity displayed. A maximally spinning 
black hole is expected to have a value of $\rm{Log(F)} = 0$, and sources that 
are not maximally spinning are expected to have values of $\rm{Log(F)} < 0$. 
There are several sources with values of $F$ greater than one, or $\rm{Log(F)}$ greater than zero. 
To see if the number of such sources is similar to that expected given the
mean uncertainty of $\delta \rm{Log(F)} = 0.15$ per source for the sources 
listed in Tables 1 and 2, consider 
dividing 
the sources into two populations: those that are maximally spinning, and thus have $\rm{Log(F)} = 0$, and a second population with some distribution of $\rm{Log(F)}$, all of which have $\rm{Log(F)}$ less than zero. 

The population of maximally spinning sources is illustrated with a 
Gaussian distribution centered on $\rm{Log(F)} = 0$, with a 
standard deviation equal to the mean uncertainty per source,
$\sigma = 0.15$, and the peak height is determined by the number of sources 
that are maximally spinning (and it is determined below that about 66 of 
the 100 sources fall into this category) so the maximum height 
of the Gaussian is $66/\sqrt{(2 \pi)}$, as illustrated with 
the dotted line in Fig. 5. This Gaussian provides a good description 
of the sources with $\rm{Log(F)} > 0$, and the properties of the 
second population can be deduced by assuming the population of 
maximally spinning holes is symmetric about $\rm{Log(F)} = 0$
and subtracting this population from the  
total number of sources. 
There are 22 
sources with $0 \leq \rm{Log(F)} \leq 0.15$, which 
indicates that for a population of sources with an intrinsic value of 
$\rm{Log(F)} =0$ and a Gaussian distribution of uncertainties we expect there to be about nine   
sources with $0.15 \leq \rm{Log(F)} \leq 0.3$ $\rm{Log(F)}$, 
and about one with $0.3 \leq \rm{Log(F)} \leq 0.45$. 
For the sample studied here,  
there are ten sources between +1 
and +2 $\sigma$ for $\rm{Log(F)} > 0$ (including the two sources 
with $\rm{Log(F)} \simeq 0.31$ with this group), 
one source between +2 
and +3 $\sigma$, and zero sources that deviate by more
than +3 $\sigma$. This is just about as expected for a population of sources centered at 
$\rm{Log(F)} =0$ with $\sigma \simeq 0.15$. 

Extending this to the sources with $\rm{Log(F)} < 0$, 
we can obtain the number of sources over and above 
that expected based on a symmetric distribution about $\rm{Log(F)} = 0$
of the population of maximally spinning black holes 
to study the 
properties of the second population of sources. 
This indicates that, over and above the sources 
expected from the Gaussian distribution (based on the numbers listed above), 
there are 13 additional sources with 
$-0.15 \leq \rm{Log(F)} < 0$, 11 additional 
sources with $-0.3 \leq \rm{Log(F)} < -0.15$, and 
ten additional sources with 
$-0.6 < \rm{Log(F)} < -0.3$, for a total of 34 sources above those 
expected from the Gaussian distribution. 
This suggests that the sources studied here consist of 
two populations: a single population of maximally spinning black holes with $\rm{Log(F)} = 0$ and 
$\sigma (\rm{Log(F)}) = 0.15$ with a total of 66 sources, plus another population 
that has a tilted distributed in $\rm{Log(F)}$, all 
of which have $-0.6 \leq \rm{Log(F)} < 0$, with a total of 34 sources. 

The number of sources 
per unit $\rm{Log(F)}$ in this second population is about 
90 for $-0.15 \leq \rm{Log(F)} < 0$, about 70 for $-0.3 \leq \rm{Log(F)} < -0.15$, and about 30 for the remainder of the sources, which have 
$-0.6 < \rm{Log(F)} < -0.3$. Part or all of this decline in the 
number of sources per unit $\rm{Log(F)}$ as $\rm{Log(F)}$ decreases 
could be due to observational selection effects, although  
part could be due to an intrinsic decline. 

These values indicate that about two-thirds of the sample of 100 
supermassive black holes are maximally spinning and 
are described by a Gaussian distribution about $\rm{Log(F) = 0}$ with 
$\sigma \simeq 0.15$. About one-third of the sample are less than 
maximally spinning and have a tilted  distribution of spin functions, with the $\rm{Log(F)}$ ranging from 
about (-0.6 to 0), with the number of sources per unit $\rm{Log(F)}$ 
declining as $\rm{Log(F)}$ decreases,  
as illustrated in Fig. 5.

The redshift distribution of $\rm{Log(F)}$ is shown in Fig. 6. 
The FRII radio sources are categorized based on their spectroscopic nuclear 
properties. The sample considered here includes high excitation galaxies (HEG),  
low excitation galaxies (LEG), quasars (Q), and weak sources (W) and each 
type is represented by a different color; 
the classifications listed here were obtained from Grimes et al. (2004).
An unweighted fit is provided, and the fitted parameters are summarized in Table 3. It is clear from Fig. 6 that sources with low values 
of $\rm{Log(F)}$ drop out of the sample as redshift increases. This 
is because sources with lower radio luminosity have lower beam power  
and thus lower values of  $\rm{Log(F)}$; the beam power is discussed in 
more detail in section 4.3 (see also D16 and D19). The radio selection effect that 
causes sources with lower radio luminosity to drop out of the sample 
as redshift increases causes sources with lower values of 
$\rm{Log(F)}$ to drop out of the sample as redshift increases.

Thus, supermassive black holes with a 
broad range of values of $\rm{Log(F)}$ are present at low redshift, while 
those with low values of $\rm{Log(F)}$ drop out as redshift increases from 
zero to two. This selection effect causes a dearth of sources with 
low values of $F$ or $\rm{Log(F)}$ at high redshift, which is clearly evident in Figs. 
5 and 6, and is due to the flux-limited nature of the survey from 
which the sources studied here are drawn.  This same selection effect 
is also apparent in all of the quantities that depend only upon $\rm{Log(F)}$.

  \begin{figure}
    \centering
    \includegraphics[width=\columnwidth]{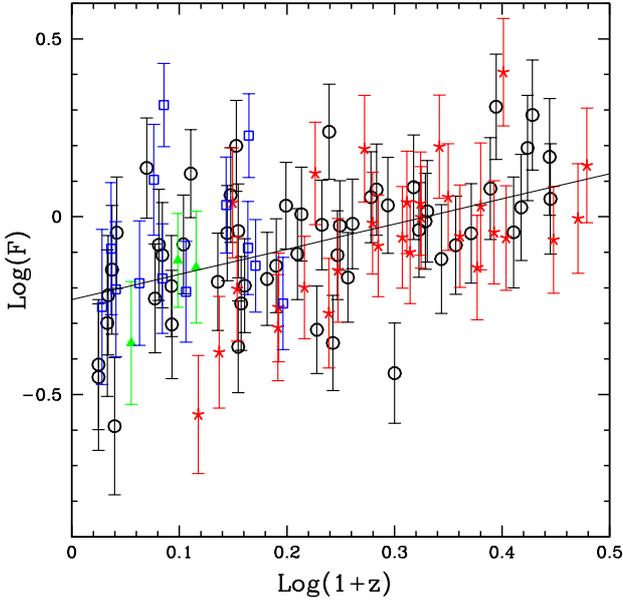}
\caption{The 
redshift (z) distribution of $\rm{Log(F)}$ is shown here. 
The theoretically 
expected maximum value of this quantity is 0. 
In this and all similar figures, HEG are denoted by open black circles, Q are denoted 
by red stars, LEG are denoted by blue squares, and W are denoted by green triangles. 
The parameters describing the best fit line in this and all similar figures are listed in 
Table 3; all fits are unweighted. This is also the redshift distribution for 
the quantity $\rm{Log(M_{rot}/M_{irr})}$ (see eq. 14).}
		  \label{fig:Fig15}
    \end{figure}

\section{Results}
The data for a sample of 100 FRII sources presented and discussed by D16 \& D19 are considered and applied here. The results are listed in Tables 1 and 2,   
and summarized in Table 3, where the typical uncertainty per source of each quantity 
is included in (brackets). 
Full details obtained with high excitation radio 
galaxies (HEG) 
are included in Table 1 while those obtained with low excitation 
galaxies (LEG), radio loud quasars (Q), and weak sources (W) 
(as defined by Grimes et al. 2004) are listed in Table 2. 
Included in Tables 1 and 2 are the $\rm{Log(F)}$ values 
obtained by D19 and the uncertainty of each value is 
also included here. 
 
 The values of $F$ listed in Tables 1 and 2 were substituted into eqs. (9-12) to solve for $\rm({M_{dyn}}/\rm{M_{irr}})$, $\rm({M_{spin}}/\rm{M_{dyn}})$, $\rm({M_{spin}}/\rm{M_{irr}})$ 
 and $\rm({E_{spin}}/\rm{E_{spin,max}})$, and the results 
are listed in  Tables 1 and 2 and illustrated in Figures 7 - 14. 
 Uncertainties of these 
 quantities are obtained using the expressions listed at the end of 
 section 2.2. Black hole masses obtained from McLure et al. (2004, 2006) and 
 listed by D19 were applied 
 using eq. (13) to obtain $\rm{M_{spin}}$; the results 
 are  
 illustrated in Figs. 15 and 16 and and listed in the Tables. 
 The total outflow energy, $\rm{E_T}$, was obtained as described by 
 O'Dea et al. (2009) (see also Leahy et al. 1989; Daly 2002), 
 and the values relative to the spin energy available for extraction, 
 $\rm({E_T/E_{spin}})$,  and relative to the black hole dynamical mass,  
 $\rm({E_T/M_{dyn}})$, are listed in the Tables and 
 illustrated in Figs. (17-20).   
Uncertainties for 
all quantities were obtained by propagating through from the original uncertainties on all quantities.  
In all of the histograms, the bin size was selected to be similar to the mean uncertainty of the quantity presented. It is helpful to consider the redshift distribution of each 
 quantity when viewing the histograms to get some perspective on 
the contributions to the histograms from sources at different redshift.   
 For many quantities of interest, 
 sources with low values drop out as the redshift increases, which 
 causes the low end of the histogram to be depleted of similar sources 
 that are likely to exist at higher redshift. 
 This can be explained by the fact that the parent population of 
 sources is 
 derived from a flux limited sample, as discussed for example    
 in section 2.3. 

\begin{figure}
    \centering
    \includegraphics[width=\columnwidth]{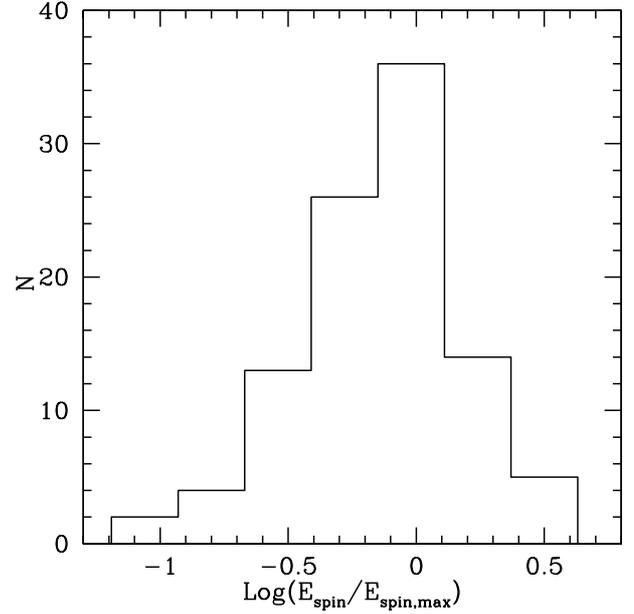}
\caption{Histogram of $\rm{Log(\rm{E_{spin}}/\rm{E_{spin,max}})}$. A  
value of $\rm{F=1}$ (i.e. $\rm{Log(F)} = 0$) substituted into eq. (12) 
indicates an expected maximum value of this quantity of 0. The sources 
with values greater than about zero are the sources with values of 
$\rm{Log(F)} > 0$. 
For more information, see the caption to Fig. 5.}
		  \label{fig:Fig4}
    \end{figure} 
    
    \begin{figure}
    \centering
    \includegraphics[width=\columnwidth]{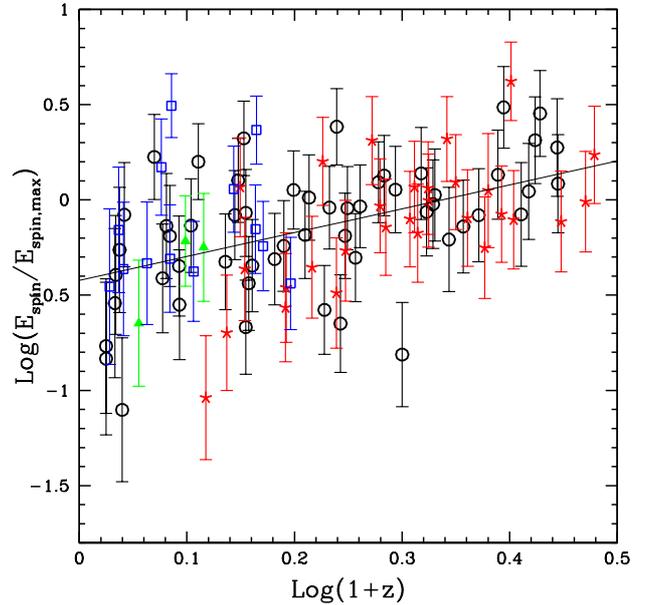}
\caption{The redshift distribution of $\rm{Log(\rm{E_{spin}}/\rm{E_{spin,max}})}$. 
A value of $\rm{F=1}$ or $\rm{Log(F)} = 0$
indicates a value of  $\rm{Log(\rm{E_{spin}}/\rm{E_{spin,max}})}$ of zero.
The symbols are as in Fig. 6 and the fit is unweighted. Values describing the 
best fit line can be deduced from those listed for $\rm{Log(M_{spin}/M_{irr}})$ in Table 3, as described in the footnote to that Table.}
		  \label{fig:Fig5}
    \end{figure} 
    
 \begin{figure}
    \centering
    \includegraphics[width=\columnwidth]{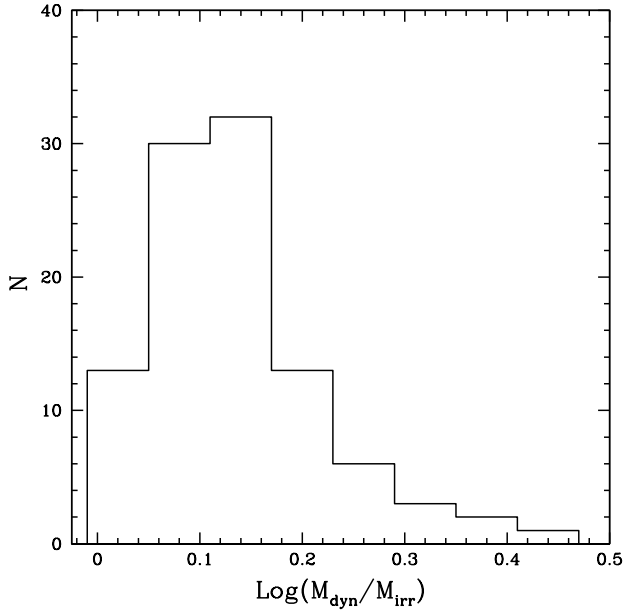}
\caption{Histogram of $\rm{Log(\rm{M_{dyn}}/\rm{M_{irr}})}$. A  
value of $\rm{F=1}$ substituted into eq. (9) 
indicates an expected maximum value of this quantity of about 0.15. The sources 
with values greater than about 0.15 are the sources with values of 
$\rm{Log(F)} > 0$; almost all of these sources have 
uncertainties $\delta \rm{Log(\rm{M_{dyn}}/\rm{M_{irr}})}$ that are within one to two sigma of 0.15. The same is true for all of the histograms that follow. 
For more information, see the caption to Fig. 5.}
		  \label{fig:Fig4}
    \end{figure} 
    
    \begin{figure}
    \centering
    \includegraphics[width=\columnwidth]{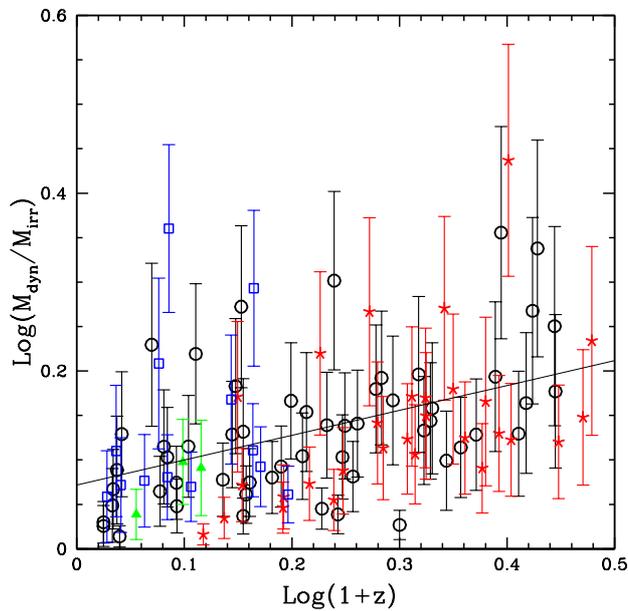}
\caption{The redshift distribution of $\rm{Log(\rm{M_{dyn}}/\rm{M_{irr}})}$. 
 A value of $\rm{F=1}$ indicates a 
value of $\rm{Log(\rm{M_{dyn}}/\rm{M_{irr}})}$ of about 0.15.
The symbols are as in Fig. 6 and the fit is unweighted.}
		  \label{fig:Fig5}
    \end{figure}    
    
    \begin{figure}
    \centering
    \includegraphics[width=\columnwidth]
    {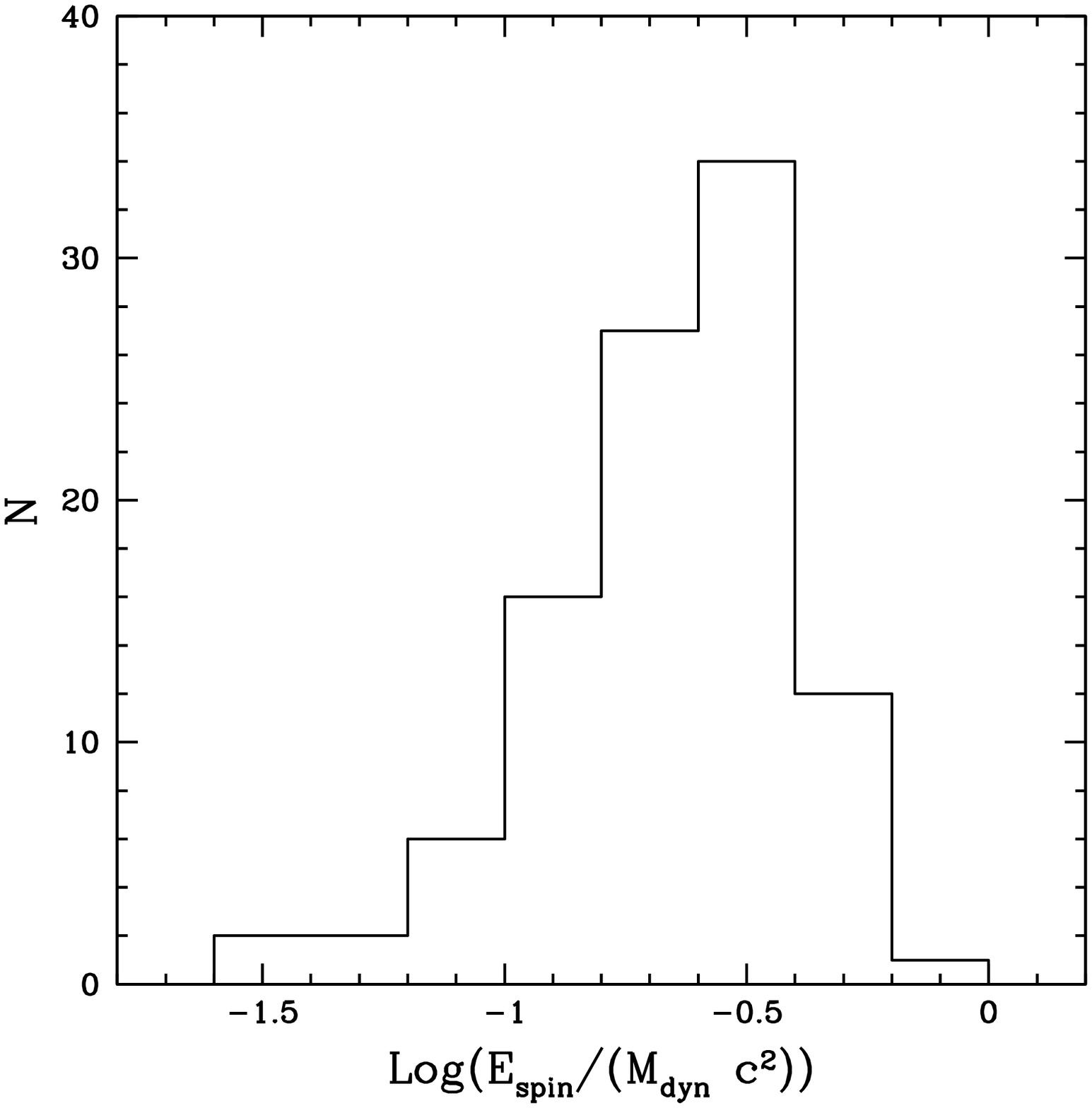}
\caption{Histogram of $\rm{Log(\rm{E_{spin}}/(\rm{M_{dyn}} c^2)}$, or $\rm{Log(\rm{M_{spin}}/\rm{M_{dyn}})}$. 
A value of $\rm{F=1}$ substituted into eq. (11) 
indicates an expected maximum value of this quantity of about -0.53. For more information, see the caption to Fig. 5.}
		  \label{fig:Fig6}
    \end{figure} 
    
   \begin{figure}
    \centering
    \includegraphics[width=\columnwidth]{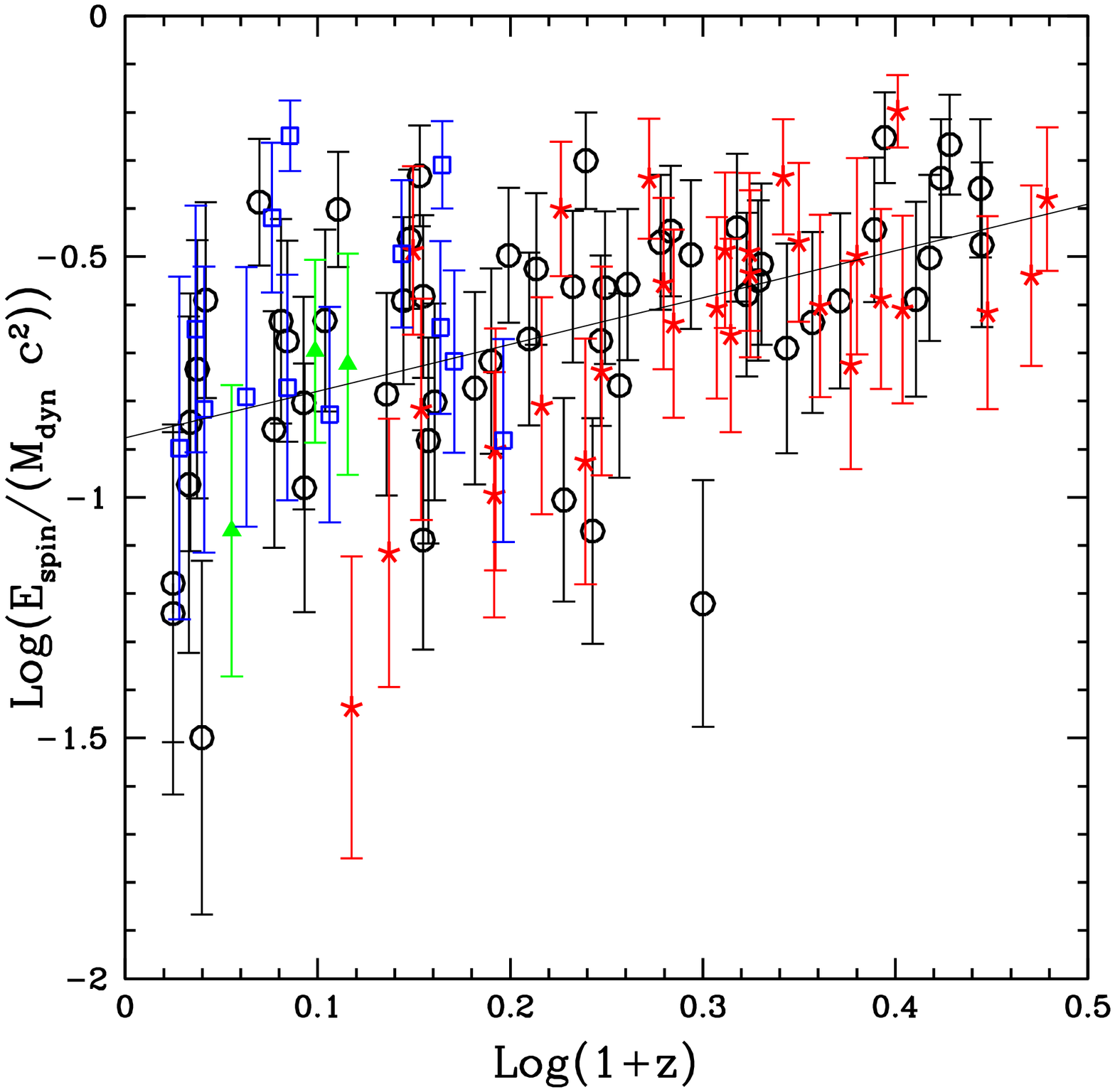}
\caption{The redshift distribution of $\rm{Log(\rm{E_{spin}}/(\rm{M_{dyn}} c^2)}$, 
or $\rm{Log(\rm{M_{spin}}/\rm{M_{dyn}})}$. The theoretically 
expected maximum value of this quantity is about -0.53. 
Symbols and information are as in Fig. 6. }
		  \label{fig:Fig7}
    \end{figure} 
      
          \begin{figure}
    \centering
    \includegraphics[width=\columnwidth]
    {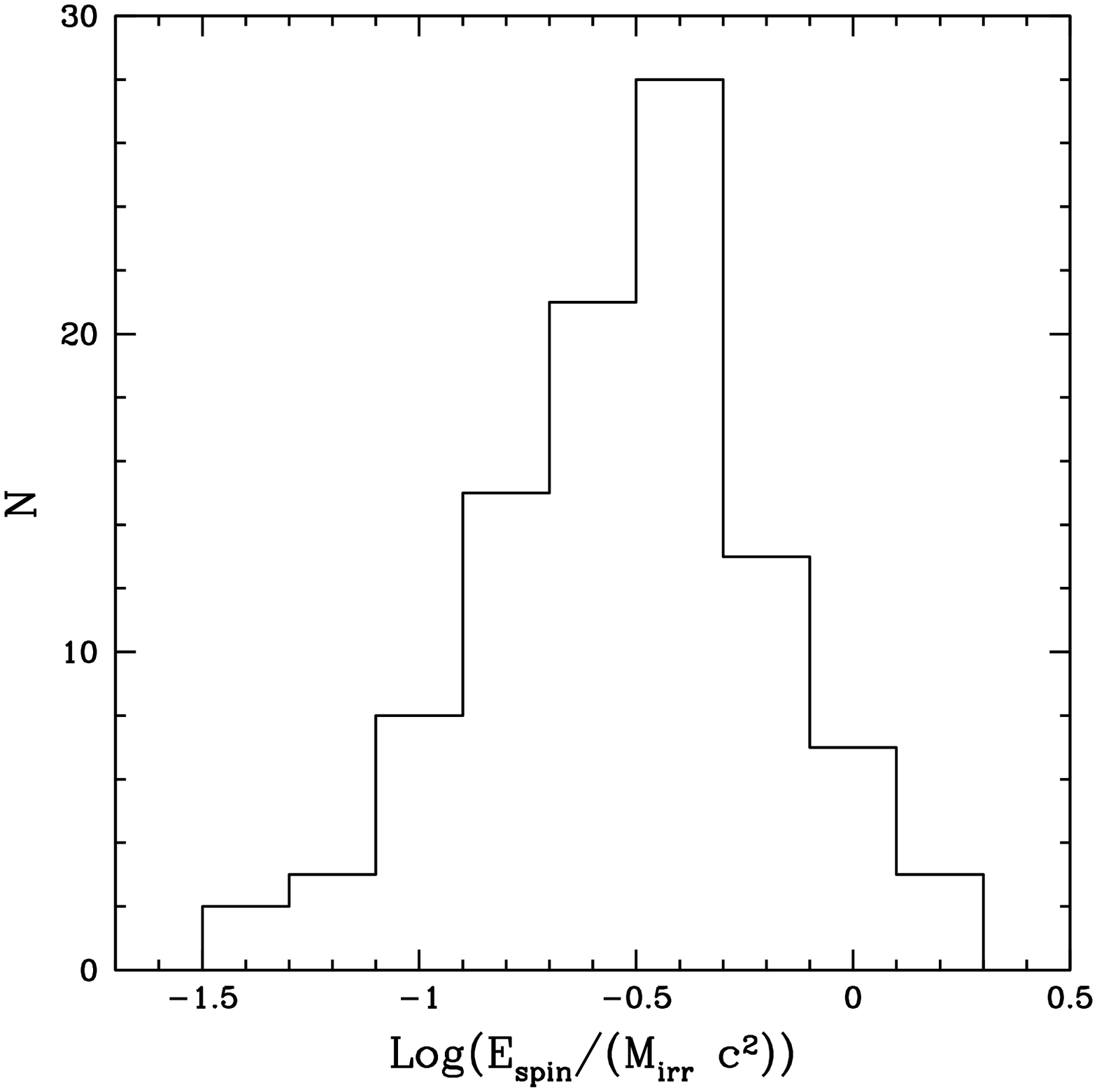}
\caption{Histogram of $\rm{Log(\rm{E_{spin}}/(\rm{M_{irr}} c^2)}$, or $\rm{Log(\rm{M_{spin}}/\rm{M_{irr}})}$. 
A value of $\rm{F=1}$ substituted into eq. (11) 
indicates an expected maximum value of this quantity of about -0.38. For more information, see the caption to Fig. 5.}
		  \label{fig:Fig8}
    \end{figure}

   \begin{figure}
    \centering
    \includegraphics[width=\columnwidth]{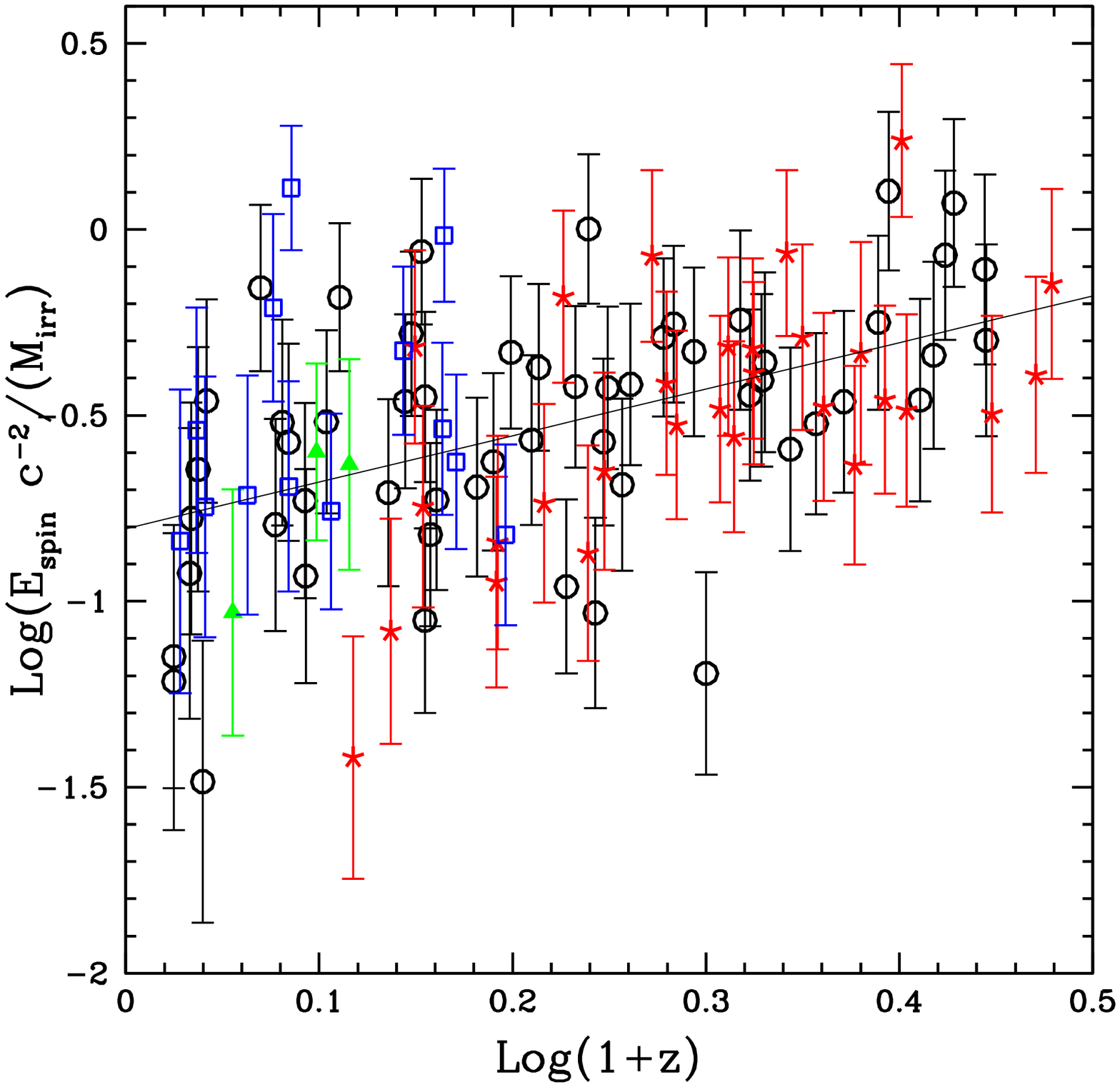}
\caption{The redshift distribution of $\rm{Log(\rm{E_{spin}}/(\rm{M_{irr}} c^2)}$, or $\rm{Log(\rm{M_{spin}}/\rm{M_{irr}})}$. 
The theoretically 
expected maximum value of this quantity is about -0.38. 
Symbols and information are as in Fig. 6.}		  
\label{fig:Fig9}
    \end{figure} 
           
\begin{figure}
    \centering
    \includegraphics[width=\columnwidth]{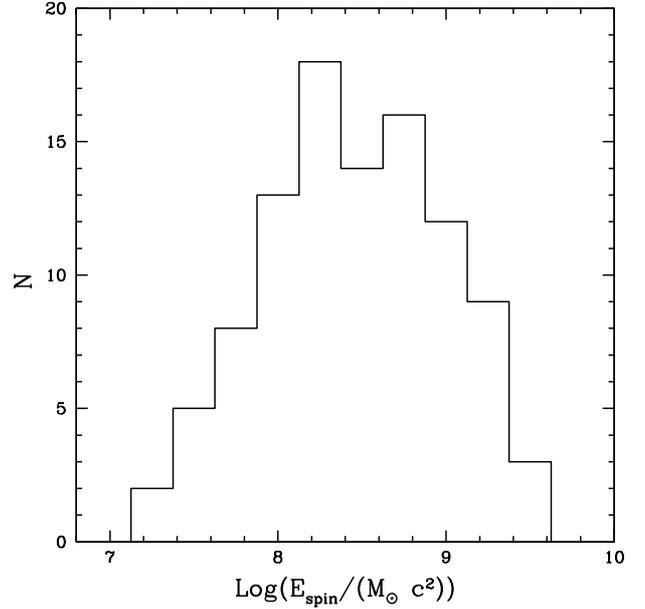}
\caption{Histogram of $\rm{Log(\rm{E_{spin}}/(M_{\odot} c^2)}$, or $\rm{Log(\rm{M_{spin}}/M_{\odot})}$ obtained with eq. (13).  
For more information, see the caption to Fig. 5. }
		  \label{fig:Fig10}
    \end{figure}

\begin{figure}
    \centering
    \includegraphics[width=\columnwidth]
    {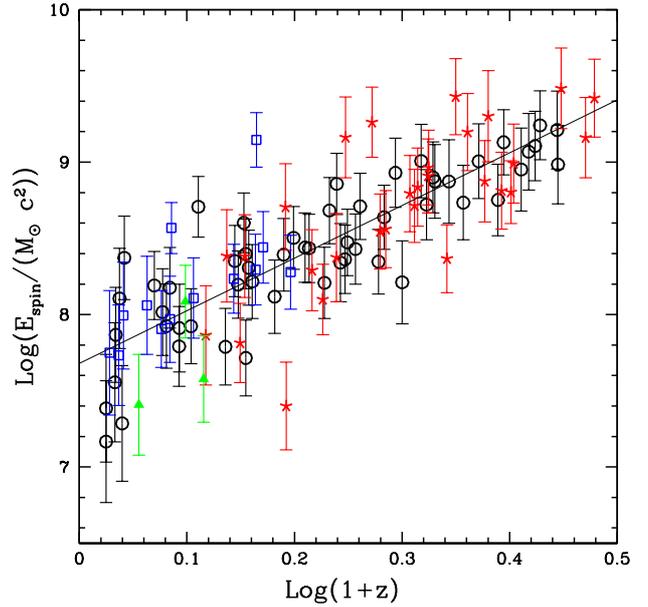}
\caption{The redshift distribution of $\rm{Log(\rm{E_{spin}}/(M_{\odot} c^2)}$, or $\rm{Log(\rm{M_{spin}}/M_{\odot})}$ obtained with eq. (13). 
Symbols and information are as in Fig. 6.}
		  \label{fig:Fig11}
    \end{figure}

\begin{figure}
    \centering
    \includegraphics[width=\columnwidth]{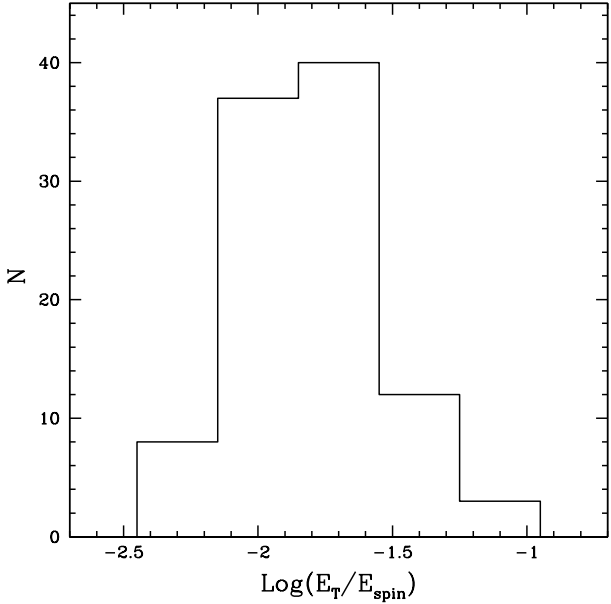}
\caption{Histogram of the Log of the total energy output by the dual collimated jets during the 
outflow event, $\rm{E_T}$, relative to the black hole spin energy available, $\rm{E_{spin}}$. 
The theoretically 
expected maximum value of this quantity is 0. 
For more information, see the caption to Fig. 5.}
\label{fig:Fig12}
\end{figure}

\begin{figure}
    \centering
    \includegraphics[width=\columnwidth]{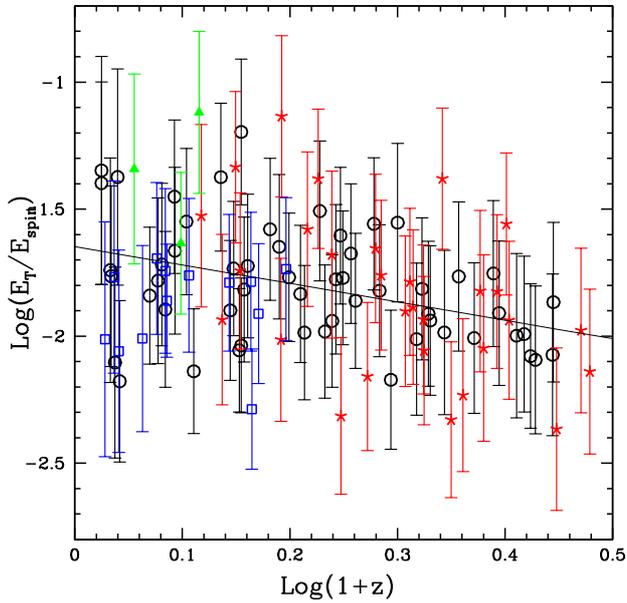}
\caption{Log of the total energy output in the form of dual collimated jets 
during the outflow event,  $\rm{E_T}$, relative to the spin energy 
available, $\rm{E_{spin}}$, vs Log(1+z). The theoretically 
expected maximum value of this quantity is 0. 
Symbols and information are as in Fig. 6.}
\label{fig:Fig13}
\end{figure}

\begin{figure}
    \centering
    \includegraphics[width=\columnwidth]{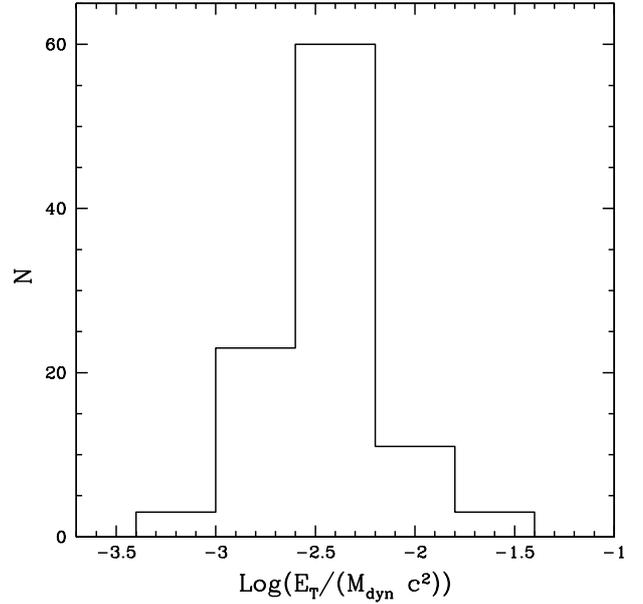}
\caption{Histogram of the Log of the total energy output in the form 
of dual collimated jets 
during the outflow event,  $\rm{E_T}$, relative to the total (dynamical) black hole 
mass, $\rm{M_{dyn}}$. For more information, see the caption to Fig. 5.}
\label{fig:Fig12}
\end{figure}

\begin{figure}
    \centering
    \includegraphics[width=\columnwidth]{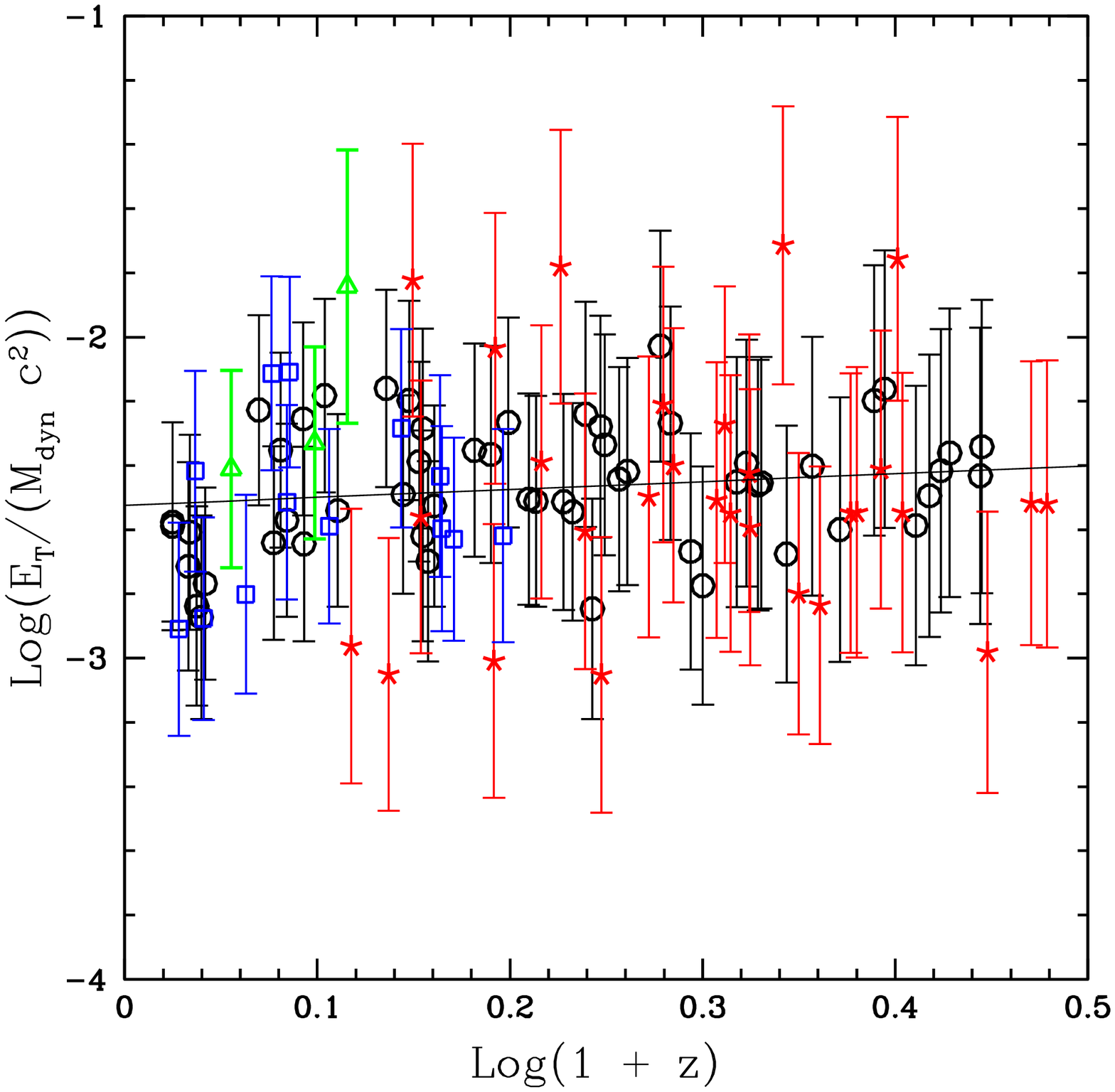}
\caption{Log of the total energy output in the form of dual collimated jets during the outflow event,  $\rm{E_T}$, relative to 
total dynamical black hole mass, $\rm{M_{dyn}}$, vs Log(1+z). 
Symbols and information are as in Fig. 6.}
\label{fig:Fig13}
\end{figure}

\section{Discussion}

\subsection{Characteristics that Depend only upon the Spin Function}

The properties of the spin function are described in section 2.3. 
The properties of the quantities obtained with the spin function reflect the 
properties of the spin function.

The fraction of the total dynamical black hole mass, $\rm{M_{dyn}}$ 
that 
is associated with the black hole spin mass-energy, $\rm{M_{spin}} 
= \rm{E_{spin}}/c^2$, typically is close to the maximum possible value 
for the classical double radio sources studied here.
For example, the mean values of HEG, Q, and LEG sources for the 
quantities $\rm{Log(\rm{M_{dyn}}/\rm{M_{irr}})}$, $\rm{Log(\rm{M_{spin}}/\rm{M_{dyn}})}$, 
$\rm{Log(\rm{M_{spin}}/\rm{M_{irr}})}$, $\rm{Log(F)}$, 
and $\rm{Log(E_{spin}/E_{spin,max}})$ 
are less than though close to the 
predicted maximum values of these quantities of about 0.15, -0.53, 
-0.38, 0, and 0, respectively (see Table 3). 
The W sources, which are all low redshift sources, 
have smaller mean values of all of these quantities relative 
to the other source types (and 
all of their values are close to the y-intercept values). This is 
not surprising since these quantities shown as a function of redshift 
clearly illustrate that sources with lower values of these 
quantities drop out as 
redshift increases due to well known selection effects. 
The classical double sources studied have redshifts between about zero and 2, 
and are selected from the 178 MHz radio flux limited
3CRR sample, described by Laing, Riley, \& Longair (1983), 
as discussed in section 2.3.
It is easy to see the impact of missing lower luminosity sources 
as redshift increases. Note that the upper envelope of the distributions 
provides a guide as to how parameters that describe sources with the largest 
spin functions, which are typically sources with the highest beam powers 
relative to the Eddington luminosity, evolve with redshift. 

The fact that black holes associated with the production of the 
classical double radio sources studied here have values of $\rm{F}$ close to 
unity and thus 
are very rapidly spinning is not 
surprising. Given that classical double radio sources are among 
the most powerful long-lived outflows observed in the 
universe, it is expected that they would be produced by  
rapidly spinning black holes with spin energies close to the maximum possible value (e.g. Rees 1984; Begelman, Blandford, \& Rees 1984; Blandford 1990). 

The values of $\rm{Log(E_{spin}/E_{spin,max})}$, 
$\rm{Log(\rm{M_{dyn}}/\rm{M_{irr}})}$, $\rm{Log(\rm{M_{spin}}/\rm{M_{dyn}})}$, 
and $\rm{Log(\rm{M_{spin}}/\rm{M_{irr}})}$ are 
listed in Tables 1 and 2 are  
consistent with or less than the maximum expected values 
about 0, 0.15, -0.53, and -0.38 within one to two sigma, 
and have distributions that reflect those of the spin functions used to obtain 
these values. 

Equation (14) indicates that the rotational mass defined in section 1 relative 
to the irreducible mass is equal to $\rm{F}$, the square root of the spin function. 
This means that the distribution of values of 
$\rm{Log(M_{rot}/M_{irr})}$ is the same as that discussed for 
$\rm{Log(F)}$ in section 2.3 for the 100 supermassive black holes studied here. 
Thus, about 2/3 (or about 66) of the 100 
sources studied here have a Gaussian distribution of $\rm{Log(M_{rot}/M_{irr})}$ 
with a mean value of zero and standard deviation of about 0.15. The remaining 1/3 
(or about 34 sources) have $\rm{Log(M_{rot}/M_{irr})} < 0$, with the tilted distribution 
described in section 2.3.  This also means that the 
values of $\rm{Log(F)}$ obtained by D19 for black holes associated with 656 additional AGN and 102 measurements of four stellar mass black holes translate directly to empirically determined values of $\rm{Log(M_{rot}/M_{irr})}$. Finally, eq. (15) indicates that values of 
$\rm{Log(M_{rot}/M_{dyn})}$ can also be obtained for the 100 sources studied here 
plus the additional AGN and stellar mass black holes mentioned above.

\subsection{Spin Mass-Energy}

The spin mass-energy per source available for extraction 
is obtained using eq. (13) where $\rm{M} = \rm{M_{dyn}}$ is the 
empirically determined dynamical mass of the black hole. 
The black hole masses listed in D16 and D19 are applied here, and
were obtained from McLure et al. (2004, 2006). 
In computing the uncertainty of the spin mass-energy that is listed 
in the Tables, the 
way that the empirically determined black hole mass enters into the 
empirically determined black hole spin function $F \propto \rm{M_{dyn}}^{-0.28}$ (e.g. D19) is taken into account. 

The spin mass-energy associated with black holes is an energy reservoir 
that is available to be tapped and when tapped can significantly affect the 
black hole environment; this is referred to as the 
"spin energy reservoir."  
For supermassive black holes, this can 
significantly affect the 
host galaxy and the environment in the vicinity of the host galaxy, 
as discussed in section 1 (see also Donahue \& Voit 2022 and references therein). 

As indicated in Figs. 15 and 16, 
the energy that is available per black hole is quite substantial. 
Since the black hole mass associated with 
classical double radio sources is strongly evolving with redshift, 
so is the spin mass-energy (see eq. 13). It is clear that sources at lower redshift 
contribute to the low mass end of the histogram while sources at 
higher redshift contribute to the high mass end of the histogram. 
The spectroscopic types that contribute to the lower spin energy end of 
the histogram include LEG and W sources, which are prevalent at 
lower redshift, while Q sources are 
prevalent at higher redshift  
and contribute preferentially to the high spin energy end of the histogram. 
The HEG sources contribute at all redshifts, as is evident from Fig. 16. 

\subsection{Total Outflow Energy Relative to Spin Mass-Energy and Relative to  Dynamical Black Hole Mass}

The fraction of the available spin energy that is produced per outflow event, 
$(\rm{E_T/E_{spin}})$, 
is obtained by dividing the total energy that is carried away 
from the black hole system during the outflow event, $\rm{E_T}$, by the 
spin energy that is available, $\rm{E_{spin}}$. 
And, the total outflow energy relative to the total 
(dynamical) black hole mass is $\rm({E_T/M_{dyn}})$. 
Note that the empirically determined quantities $\rm{E_T}$ and 
$\rm{M_{dyn}}$ are obtained with completely independent methods.  
The range of values for the total outflow energy per source, 
$\rm{E_T}$, span about an order of magnitude 
(e.g. see Figs. 40 and 41 from O'Dea et al. 2009), the range of  
values of $\rm{E_{spin}}$ span about two orders of magnitude 
(see Figs. 15 and 16), and the range of values of values of 
$\rm{M_{dyn}}$ span about two orders of magnitude 
(see Fig. 3 of D19). 

The total outflow energy is obtained by multiplying the total 
outflow timescale by the beam power, where the beam power is the 
energy per unit time output in the form of dual jets from the 
black hole system (e.g. O'Dea et al. 2009). It has been shown 
conclusively for classical double (FRII) sources such as those 
studied here that the total outflow 
timescale is very well characterized as a function of only the beam power 
(Daly 1994; Daly et al. 2008, 2009). 
Note that the relationship between the total outflow timescale 
and the beam power is the foundation of the use of classical 
double radio galaxies for cosmological studies. The fact that 
this application for cosmological studies yields results that 
are very similar to and consistent with those obtained with other methods indicates that 
this model is on secure footing, as discussed in detail, 
for example, by Daly et al. (2008, 2009). 

The total outflow energy per source obtained by O'Dea et al. (2009) 
is used here, and an identical method is applied to obtain the total 
outflow lifetime from the beam power and 
thus the total outflow energy for 
the remaining sources in the sample. 
The total outflow energy per source, referred to as 
$\rm{E_T}$, is divided by the spin energy $\rm{E_{spin}}$ to 
obtain the fraction of the spin energy that could be extracted per outflow 
event, $(\rm{E_T/E_{spin}})$. And, $\rm{E_T}$ is divided by the black hole mass 
$\rm{M_{dyn}}$ to obtain the fraction of the black hole mass 
that is produced per outflow event, $(\rm{E_T/M_{dyn}})$.
Note that the total outflow energy $\rm{E_T} $ is independent of the black hole 
mass and only depends on the beam power of the source, which is 
empirically determined 
using the strong shock method (reviewed in detail by 
O'Dea et al. 2009). 

The results obtained here indicate that only a small fraction, 
about 1.5\% 
of the spin energy available per black hole is produced per outflow event; see the values listed in Tables 1 and 2, and summarized 
in Table 3. The fraction $\rm{(E_T/E_{spin})}$ is independent of source type (see Table 3), except 
for the W sources, and there are only three low redshift W sources in the sample.
The results indicate that the mean value of $\rm{Log(E_T/E_{spin})}$ for the 100 sources studied is about 
$\rm{Log(E_T/E_{spin})} \simeq -1.81 \pm 0.26$. This translates to a small fraction of the black hole 
dynamical mass being output per outflow event, as indicated 
by the values of $\rm{Log(E_T/M_{dyn})}$ listed in Tables 1 and 2 
and summarized in Table 3. The mean value of 
this quantity is $\rm{Log(E_T/M_{dyn})} \simeq -2.47 \pm 0.27$ for the 100 sources studied. This translates to a mean value of the total 
outflow energy relative to dynamical black hole mass of about 
$\rm{(E_T/M_{dyn}}) \simeq 3.4 \times 10^{-3}$.
These results are 
consistent with those obtained by Daly (2009a) who 
studied a sample of 19 classical double radio sources and 
found that about a few $\times 10^{-3}$ of the black hole dynamical mass 
is output in the form of large-scale jets per source per 
outflow event. As mentioned earlier, there is no overlap in the 
methods used to obtain $\rm{E_T}$ and $\rm{M_{dyn}}$.

There are several possible explanations for the fact that 
the total energy output over the source lifetime in the form of large-scale jets is small compared with the black hole dynamical mass and compared with the 
spin energy available for extraction, and that each 
has a relatively narrow distribution.
1. When a certain fraction of the black hole mass-energy is 
deposited into the ambient gas, the gas is heated and expands, and 
the accretion is shut off; this would be consistent with the result 
obtained here and by Daly (2009a).
2. The spin energy extraction, which decreases the black hole 
dynamical mass, destabilizes the black hole - accretion disk -magnetic field configuration causing the spin energy extraction to be terminated. 
3. The black hole masses have been overestimated, and the total 
spin energy available for extraction is smaller than obtained based on current 
black hole mass estimates; this would increase the ratio $\rm{(E_T/E_{spin})}$ and the ratio $\rm{(E_T/M_{dyn})}$. 
4. The beam powers are much much larger 
than indicated empirically, and thus carry away significantly 
more energy than already accounted for. 5. The black hole spin 
function $F$, and thus dimensionless spin angular momentum and 
spin energy, has been overestimated. 
This would only impact $\rm{E_{spin}}$ and thus $\rm{(E_T/E_{spin}})$, 
but would not impact $\rm{M_{dyn}}$ and thus would not impact 
$\rm{(E_T/M_{dyn})}$.
6.  
Only transitions between particular spin states 
are allowed, as described by Pugliese \& Quevedo (2022) and Pugliese \& Stuchlík (2021). 
7. Something else. Each of these possibilities is considered. 

Possibility 1. could explain the observed values and small range of values  
of the quantities $\rm{(E_T/E_{spin})}$ and $\rm{(E_T/M_{dyn})}$ obtained here and 
by Daly (2009a). 
The results indicate that the energy 
deposited into the ambient gas over the entire lifetime of an FRII 
source relative to the black hole dynamical mass is about 
$\rm{Log(E_T/M_{dyn})} \simeq (-2.47 \pm 0.27)$ 
(see Table 3 and Fig. 19 in this work,  and Table 1 and Fig. 1 from Daly 2009a). 
These results are consistent with  
the empirically determined value of about 
$-2.3 \pm 0.5$ obtained by Donahue \& Voit (2022) (see their Fig. 20) 
based on empirical studies of the energy input required to heat and 
lift the 
circumgalactic medium and shut off accretion  
for a sample of relatively low redshift sources. One interesting caveat 
is that the FRII sources studied here have redshifts between about zero 
and two, and the source sizes change   
significantly with redshift (e.g. 
Fig. 8 of Guerra, Daly, \& Wan 2000), so the result obtained here 
would have to be independent of the details of the energy input such as 
where in the galactic and circumgalactic medium the 
energy is deposited and independent of the structure (density and temperature) 
of the galactic and circumgalactic medium.  
 
In this scenario, the 
accretion would be shut off by the heating and lifting of the circumgalactic medium; the medium would eventually settle down and another outflow episode would occur. Each outflow event would decrease the 
black hole spin energy by a very small amount, as long as the angular momentum extracted during the outflow event exceeds that gained by the black hole during the accretion event. One puzzling factor for this 
interpretation is that the range of values of $\rm{(E_T/M_{dyn})}$ 
and $\rm{(E_T/E_{spin})}$ 
obtained here and by Daly (2009a) are narrow, and seemingly independent 
of radio source size and source redshift (see Fig. 20, and the value of the 
slope listed in Table 3). 

One rather radical idea to explain the small values and small range  
of these quantities is to posit that the 
majority of the spin energy is extracted per outflow event, but most of it does not end up
in the form of a dual collimated outflow (which would comprise a set 
fraction of the total energy extracted per unit time), but is in 
some difficult to detect form such as neutrinos,  
or gravitational waves. In the outflow method, the normalization of eq. (1) is a free parameter 
that is empirically determined. The empirically determined value is consistent with 
the theoretical prediction in the Meier (1999) model 
(see section 3.3 of D19), and is also consistent with the 
normalization in the Blandford \& Znajek (1977) model. 

Thus, this hypothetical other process would occur 
simultaneously with the Blandford \& Znajek (1977) or Meier (1999)  
mechanism but would extract substantially more spin energy per unit time, by factors of about (10 - 100), 
and 
the energy extracted would be in some form that is not readily observable. This process could work hand-in-hand with possibilities 
2 and/or 3. 

Note that for FRII sources the outflow timescale 
depends only upon the beam power, 
indicating that the accretion timescale must exceed the outflow timescale 
unless some process directly related to the beam power shuts off the accretion.
Otherwise, the outflow timescale would be set by the accretion 
timescale and would not be a function of only the beam power, 
as has been shown conclusively by Daly et al. (2009). 

Possibility 2. is quite interesting. As the spin energy is extracted, 
the black hole mass decreases causing the accretion disk to expand slightly and over a long period of time; the outflow timescales 
are typically a few $\times 10^7$ years (e.g. O'Dea et al. 2009). 
If the stability of the magnetic field that plays a crucial role in the 
spin energy extraction requires a particular ratio of the disk thickness 
to the disk radius, as the disk expands the thin disk may be disrupted.
That is, it is possible that the disk and thus the anchor of the magnetic field is disrupted when the fraction 
of the black hole dynamical mass is decreased by the particular value 
of a few tenths of a percent found here and 
by Daly (2009a). The decrease of 
the black hole mass would have a small impact on the radius of the 
disk, but could have a large impact on the disk thickness, which 
is likely to be small relative to the disk radius 
(see, for example, Blandford \& Globus 2022; 
Kolos et al. 2021). 
Possibility 2. could work hand-in-hand with possibility 1. 
It's not clear how large a fraction of the black hole mass-energy 
would have to be removed to de-stabilize the accretion disk - magnetic field - black hole configuration and thus terminate the outflow. This 
possibility would be more palatable if the fraction of the black hole 
mass removed was larger, as considered in point 3. 

This brings us to possibility 3. If the black hole masses have 
been systematically overestimated, then the spin energy values obtained with eq. (13) 
decrease and the ratios $\rm{(E_T/E_{spin})}$ and $\rm{(E_T/M_{dyn})}$ increase. 
There are some recent studies that suggest that black hole 
dynamical masses may be systematically overestimated 
(e.g. Grier et al. 2019). However, the brightest sources 
studied here and by D19
have 
a bolometric accretion disk luminosity that is right at the 
Eddington luminosity (see Fig. 4 of D19), and any decrease in black hole 
mass would cause these sources to be radiating at super-Eddington levels. 

Possibility 4 is very unlikely based on the following. 
The direct comparison between the total outflow energy and 
the black hole mass indicates that the outflow energy is 
a roughly constant fraction of the black hole mass, 
with $\rm{(E_T} /\rm{M_{dyn})} \approx 3 \times 10^{-3}$ independent of the 
spin properties of the black hole (see Figs. 19 and 20, 
Tables 1, 2, and 3, and 
Daly 2009a). As noted by O'Dea 
et al. (2009), the total outflow energy scales as the beam power 
$L_j^{0.5}$, so to significantly increase the outflow 
energy by factors of 10 to 100, the beam power would have to 
increase by factors of $10^2 - 10^4$, which is highly unlikely 
since 
the beam power is insensitive to offsets from minimum energy conditions (e.g. O'Dea et al. 2009). In addition, 
the largest beam powers are about 10 \% of the Eddington luminosity 
(e.g. Daly et al. 2018), so this would require the maximum beam powers to be significantly larger than the Eddington luminosity. And, as noted above, 
the empirically determined beam power normalizations match those predicted 
theoretically in the Meier (1999) and Blandford \& Znajek (1977) models. 

Possibility 5. is unlikely because independent spin determinations 
for supermassive 
black holes associated with classical double radio sources agree with 
those obtained with the outflow method, and indicate high spin values
(e.g. Azadi et al. 2020). Fifteen of the quasars studied by D19 with the 
outflow method 
overlap with those studied by Azadi et al. (2020) with the continuum fitting 
method, and the spin values obtained with the independent methods agree. 
Similarly, for local AGN, spin values obtained 
with the outflow method agree with those obtained independently with 
the X-ray reflection method for the six sources for which a comparison was possible (D19). 
Possibility 5. would require that spin determinations published to date for supermassive black holes 
by other groups using independent 
methods are incorrect by large factors. 

Other options are possibility 6., 
only transitions between particular black hole 
spin states are allowed as described by Pugliese \& Quevedo (2022) and Pugliese \& Stuchlík (2021), or possibility 7, something else.

\section{Summary}

Mass-energy characteristics of black holes are obtained 
in terms of the black hole spin function, $\rm{F^2}$. Empirically determined 
black hole spin functions are used to obtain and 
study the spin mass-energy properties of a sample of 100 supermassive 
black holes associated with classical double (FRII) radio sources with dual 
collimated outflows; the sources have redshifts between about zero and two. 
Black hole spin mass-energy that is 
available to be extracted from the black hole is $\rm{M_{spin} = M- M_{irr}}$, where  $\rm{M \equiv M_{dyn}}$ 
(see eq. 2). The mass-energy associated with the 
black hole spin angular momentum $\rm{J}$, referred to here as 
$\rm{M_{rot}}$ and defined in section 1, contributes to the total black hole 
mass, M:  $\rm{M^2 = M_{irr}^2 +M_{rot}^2}$, which leads to eqs. (3) and (9). 
These equations are combined to obtain 
expressions that describe black hole spin mass-energy characteristics 
in terms of the spin function, which are then  
applied to quantify and study 
empirically determined black hole spin mass-energy properties for a sample of 100 supermassive black holes. It is important to be able to empirically determine 
black hole spin mass-energy characteristics because these impact 
the total black hole 
mass, and because this energy can be extracted, which may impact 
the near and far field environments of astrophysical black holes.

The relationship between the beam power in Eddington units and 
bolometric accretion disk luminosity in Eddington units for the 
sample of supermassive black holes studied here is very similar to 
and consistent with that obtained for three other samples of sources 
with very different ranges and values of Eddington normalized beam power and bolometric 
disk luminosity (Daly et al. 2018). The samples studied include 
the 100 sources studied here plus 656 AGN and 102 measurements of four stellar-mass 
black holes that are in X-ray binary systems, and include several different types of AGN. 
This suggests that the outflows in all of these 
systems are produced by a common physical mechanism. Since many of the sources 
studied by Daly et al. (2018) have beam powers that are much larger (by factors 
of 10 to 100) than the 
bolometric accretion disk luminosity, these sources are likely to have spin-powered outflows. 
Since the outflows in all of the sources studied are likely to be produced by a common 
physical mechanism, this suggests that all of the sources, including those studied here, 
have spin powered outflows. 

Quantities that characterize the spin mass-energy properties of 
astrophysical black holes in terms of the black hole
spin function, $\rm{F^2}$, are presented in section 2.2. This is preferable 
for astrophysical black holes for several reasons. 
For example, when attempting 
to use the dimensionless black hole spin angular momentum $\rm{j \equiv Jc/(G M^2)}$ 
to empirically characterize and determine the spin properties of 
astrophysical black holes, several difficulties are encountered, 
as described in section 2.1. 
These issues may be avoided and circumvented by 
writing the black hole spin mass-energy characteristics 
in terms of the 
black hole spin function $\rm{F^2}$.
Furthermore, in the context of the outflow method, the empirically determined 
quantity is $\rm{F}$. 

Relationships between the black hole 
spin mass-energy characteristics and the black hole spin function 
$\rm{F^2}$ are obtained and presented in section 2.2. 
It is found that there is roughly 
a linear relationship between the black hole spin function and 
the normalized 
spin mass-energy of the black hole $\rm{(E_{spin}/E_{spin,max}) \approx F^2}$, and allowing the exponent of $\rm{F}$ to vary, that 
$\rm{Log(E_{spin}/E_{spin,max}) \approx 1.75~ Log(F)}$ 
over the range of values 
relevant to the current studies. In addition, the method 
allows for empirically determined values of the spin function that 
exceed unity, which can occur due to the uncertainties associated with 
empirically determined quantities for astrophysical black holes. 

The method described in section 2.2 
is applied to a sample of 100 supermassive black holes 
with redshifts between about zero and 2. 
The values of $\rm{Log(F)}$ studied here were obtained by D19, and 
are listed along with their uncertainties in Tables 1 and 2. It is shown 
in section 2.3 that the sample is well represented as having two components: 
about 2/3 of the 100 sources are maximally spinning, and about 1/3 are less 
than maximally spinning with the number of sources per unit $\rm{Log(F)}$ 
declining as  $\rm{Log(F)}$ decreases. The decreasing number of sources as 
 $\rm{Log(F)}$ decreases could be due to observational selection effects, 
 a real decline with  $\rm{Log(F)}$, or a combination of the two. 
 The 100 FRII sources studied include four sub-samples based on their 
 spectroscopic nuclear properties; HEG, LEG, Q, and W sources, as described in section 2.3. As is evident from Table 3, the 
 results presented here are, for the most part, independent of source 
 spectroscopic nuclear properties, except for the W sources, and there 
 are only three low-redshift W sources in the sample. 
 
 Interestingly, 
 it turns out that $\rm{Log(M_{rot}/M_{irr}) = Log(F)}$ (see eq. 14), 
 so all of the comments and results obtained for 
 $\rm{Log(F)}$ directly apply to $\rm{Log(M_{rot}/M_{irr})}$. Thus, the 
 distribution of values of $\rm{Log(F)}$ described in section 2.3 
 can be interpreted as the empirically determined 
 distribution of values of $\rm{Log(M_{rot}/M_{irr})}$. The 
 empirically determined values of $\rm{Log(F)}$ and their 
 uncertainties for an additional 
 656 AGN and 102 measurements of four stellar mass black holes listed 
 and discussed by D19 also directly translate to values of $\rm{Log(M_{rot}/M_{irr})}$ for those sources. 
 The quantity 
 $\rm{Log(M_{rot}/M_{dyn})}$ can be obtained from eq. (15), which 
 indicates that $\rm{Log(M_{rot}/M_{dyn})} = Log(F) - Log(M_{dyn}/M_{irr})$, 
 both of which are listed in Tables 1 and 2.

Results describing the spin mass-energy characteristics 
of the 100 sources 
are presented and discussed in sections 3 and 4. Many of 
the sources are highly spinning, and the sources with lower values of 
black hole spin are at low redshift, as expected due to the flux 
limited nature of the parent population of the sources. 
Thus, the fact that many of the sources are highly spinning may 
be a selection effect in that the most highly spinning sources have the 
brightest and most powerful radio emission, and less powerful sources 
drop out of the sample at high redshift due to the flux limited 
nature of the parent population, as described in sec. 2.3.

The spin mass-energy values obtained from the black hole spin functions 
are studied relative to the total 
or dynamical black hole mass and relative to the irreducible 
black hole mass. For maximally spinning black holes, the 
mass-energy associated with the black hole spin contributes about 
41 \% relative to the irreducible black hole mass or about 
29 \% relative to 
the total dynamical black hole mass. This mass-energy can be extracted 
(Penrose 1969). 
Thus, the mass of the black hole can be decreased due to the extraction of 
the spin energy. In addition, the extraction of the spin energy can significantly affect 
the short and long range environment of each black hole. Since these are 
all FRII (classical double) radio sources, these sources channel energy significant 
distances (hundreds of kpc) from the supermassive black hole.

The spin mass-energy relative to the 
dynamical (i.e. total) black hole mass can be combined 
with empirical determinations of the black hole mass to solve for the 
total spin energy available for extraction per source, as 
discussed in detail in sections 1 and 2 (see eq. 13). 
The spin energy per supermassive black hole 
is substantial, 
and represents an important reservoir of energy that can be tapped; this is
referred to as the "spin energy reservoir." Tapping even small amounts of the 
spin energy can have a substantial impact on the near and far field environments 
of the sources, as discussed in sections 4.2 and 4.3.

The total spin energy available per source is 
compared with the total energy output from the black hole system 
in the form of dual oppositely directed jets over the active lifetime of each 
source, $\rm{E_T}$, as described in sections 3 and 4.3. 
For the 100 black hole systems studied, the range of values of 
$(\rm{E_T} /\rm{E_{spin}})$, the ratio of the total outflow energy  
to the spin energy 
available, is very narrow, with most of the sources having a value of 
about one percent or so: $\rm{Log(E_T/E_{spin})} \simeq 
-1.8 \pm 0.3$ for the 100 FRII sources studied here. 
This is consistent with the results obtained 
here and by 
Daly (2009a) that indicated a small value and 
range of values of total outflow energy 
relative to black hole dynamical mass:
$\rm{Log(E_T/M_{dyn})} \simeq 
-2.5 \pm 0.3$ for the 100 FRII sources studied here (see sections 3 and 4.3). The value obtained here is consistent with that obtained by Daly (2009a) and 
that with obtained with a different method applied to different 
types of sources by 
Donahue \& Voit (2022), who find  
$\rm{Log(E_T/M_{dyn})} \simeq 
-2.3 \pm 0.5$ for a sample of low redshift sources. 
The small value and restricted range of values of $\rm{Log(E_T/M_{dyn})}$ could suggest that this is a fundamental property of the primary process responsible for producing the dual collimated outflows.

Several possible explanations for the relatively small value 
and range of values of $(\rm{E_T/M_{dyn}})$
or $(\rm{E_T} /\rm{E_{spin}})$ are considered 
in section 4.3. For example, it could be that when a specific amount of energy 
relative to the dynamical black hole mass is dumped into the 
ambient medium, the ambient gas is heated and expands, shutting 
off the accretion. Another possibility is that as the spin energy 
is extracted and the black hole mass decreases, the magnetic field 
and/or the structure of the accretion disk is altered and the spin 
energy extraction is halted. Or, it could be that much of the spin 
energy is extracted and then the process shuts down - if the black hole 
masses have been systematically overestimated, then the black hole
mass that enters into eq. (13) is decreased and the spin energies decrease, 
so a correspondingly 
larger fraction of the spin energy is extracted per outflow event. 
Another possibility discussed in section 4.3 
is that there is some other process that 
occurs simultaneously with the process that leads to dual large-scale jets, 
and this other process is extracting the majority of the 
spin energy, but the extracted energy is released  
in a form that is not readily observable.   
For example, most of the spin energy could be carried away in the 
form of neutrinos or gravitational waves,  
and only a small fraction of the energy extracted 
would be channeled into the jetted dual outflow.

The new method of obtaining black hole spin mass-energy characteristics 
directly from the spin function presented here is applicable to the study of 
astrophysical black holes in a broad range of contexts. 

\section*{Acknowledgments}
Thanks are extended to the referee, Kastytis Zubovas, 
for a careful reading of the 
manuscript and for providing very helpful comments and suggestions.
It is a pleasure to thank Megan Donahue, Jim Pringle, and Mark Voit for 
detailed discussions and suggestions related to this work. I would also like to thank 
Jean Brodie,  
Margaret Daly, Joshua Deal, 
Yan-Fei Jiang, Chiara Mingarelli, Chris O'Dea, 
Masha Okounkova,  
Enrico Ramirez-Ruiz, Biny Sebastian, and 
Rosie Wyse for helpful conversations related to this work. 
It is a pleasure to thank the Center for Computational 
Astrophysics and the Flatiron Institute, which is supported by the 
Simons Foundation, for their hospitality. 
This work was performed in part at the Aspen Center for Physics, 
which is supported by National Science Foundation grant PHY-1607611. 

\section{DATA AVAILABILITY STATEMENT}
The data underlying this article are available in the article or are listed in D19.




\newpage

\begin{table*}																																		
\begin{minipage}{165mm}																																		
\scriptsize																																		
\caption{Outflow and Spin Properties for FRII LEG, Q, \& W Sources}																																		
\label{FRIIsources-LEG-plus W sources new-paper}																																		
\begin{tabular}{lllllllllll}   
\hline\hline  
(1)&(2)&(3)&(4)&(5)&(6)&(7)&(8)&(9)&(10)&(11)\\	
Source&Type&z&$\rm{Log}$ &	$\rm{Log}$&																																	
$\rm{Log}$&$\rm{Log}$&$\rm{Log}$&$\rm{Log(\rm{M_{spin}})}$&$\rm{Log}$&$\rm{Log}$\\																																		
&&&$\rm{(F)}$&$\rm{(\rm{E_{spin}}/E_{s,max})}$&
$\rm{(\rm{M_{dyn}}/\rm{M_{irr}})}$&$\rm{(\rm{M_{spin}}/\rm{M_{dyn}})}$&$\rm{(\rm{M_{spin}}/\rm{M_{irr}})}$&${(M_{\odot})}$&$\rm{(E_T/\rm{E_{spin}})}$&$\rm{(E_T/M_{dyn})}$\\																				
\hline
3C	33	&	HEG	&$	0.059	$&$	-0.42	\pm	0.18	$&$	-0.77	\pm	0.35	$&$	0.03	\pm	0.02	$&$	-1.18	\pm	0.33	$&$	-1.15	\pm	0.35	$&$	7.38	\pm	0.12	$&$	-1.40	\pm	0.22	$&$	-2.58	\pm	0.31	$\\
3C	192	&	HEG	&$	0.059	$&$	-0.45	\pm	0.21	$&$	-0.83	\pm	0.40	$&$	0.03	\pm	0.02	$&$	-1.24	\pm	0.38	$&$	-1.22	\pm	0.40	$&$	7.17	\pm	0.12	$&$	-1.35	\pm	0.24	$&$	-2.59	\pm	0.32	$\\
3C	285	&	HEG	&$	0.079	$&$	-0.30	\pm	0.21	$&$	-0.54	\pm	0.39	$&$	0.05	\pm	0.04	$&$	-0.97	\pm	0.35	$&$	-0.92	\pm	0.39	$&$	7.56	\pm	0.13	$&$	-1.74	\pm	0.24	$&$	-2.71	\pm	0.32	$\\
3C	452	&	HEG	&$	0.081	$&$	-0.22	\pm	0.17	$&$	-0.39	\pm	0.31	$&$	0.07	\pm	0.05	$&$	-0.84	\pm	0.27	$&$	-0.78	\pm	0.31	$&$	7.87	\pm	0.14	$&$	-1.76	\pm	0.22	$&$	-2.61	\pm	0.30	$\\
3C	388	&	HEG	&$	0.09	$&$	-0.15	\pm	0.18	$&$	-0.26	\pm	0.33	$&$	0.09	\pm	0.06	$&$	-0.73	\pm	0.27	$&$	-0.65	\pm	0.33	$&$	8.11	\pm	0.15	$&$	-2.10	\pm	0.24	$&$	-2.84	\pm	0.31	$\\
3C	321	&	HEG	&$	0.096	$&$	-0.59	\pm	0.19	$&$	-1.10	\pm	0.38	$&$	0.01	\pm	0.01	$&$	-1.50	\pm	0.37	$&$	-1.49	\pm	0.38	$&$	7.29	\pm	0.12	$&$	-1.37	\pm	0.23	$&$	-2.87	\pm	0.32	$\\
3C	433	&	HEG	&$	0.101	$&$	-0.05	\pm	0.16	$&$	-0.08	\pm	0.27	$&$	0.13	\pm	0.07	$&$	-0.59	\pm	0.20	$&$	-0.46	\pm	0.27	$&$	8.37	\pm	0.16	$&$	-2.18	\pm	0.23	$&$	-2.77	\pm	0.30	$\\
3C	20	&	HEG	&$	0.174	$&$	0.14	\pm	0.14	$&$	0.23	\pm	0.22	$&$	0.23	\pm	0.09	$&$	-0.39	\pm	0.13	$&$	-0.16	\pm	0.22	$&$	8.19	\pm	0.19	$&$	-1.84	\pm	0.24	$&$	-2.23	\pm	0.30	$\\
3C	28	&	HEG	&$	0.195	$&$	-0.23	\pm	0.15	$&$	-0.41	\pm	0.29	$&$	0.07	\pm	0.04	$&$	-0.86	\pm	0.25	$&$	-0.79	\pm	0.29	$&$	8.02	\pm	0.14	$&$	-1.78	\pm	0.21	$&$	-2.64	\pm	0.30	$\\
3C	349	&	HEG	&$	0.205	$&$	-0.08	\pm	0.16	$&$	-0.14	\pm	0.28	$&$	0.12	\pm	0.06	$&$	-0.63	\pm	0.21	$&$	-0.52	\pm	0.28	$&$	7.92	\pm	0.16	$&$	-1.72	\pm	0.23	$&$	-2.35	\pm	0.30	$\\
3C	436	&	HEG	&$	0.214	$&$	-0.11	\pm	0.15	$&$	-0.19	\pm	0.27	$&$	0.10	\pm	0.06	$&$	-0.68	\pm	0.21	$&$	-0.57	\pm	0.26	$&$	8.18	\pm	0.16	$&$	-1.90	\pm	0.22	$&$	-2.57	\pm	0.30	$\\
3C	171	&	HEG	&$	0.238	$&$	-0.20	\pm	0.14	$&$	-0.35	\pm	0.26	$&$	0.07	\pm	0.04	$&$	-0.80	\pm	0.22	$&$	-0.73	\pm	0.26	$&$	7.79	\pm	0.15	$&$	-1.45	\pm	0.21	$&$	-2.25	\pm	0.30	$\\
3C	284	&	HEG	&$	0.239	$&$	-0.30	\pm	0.15	$&$	-0.55	\pm	0.29	$&$	0.05	\pm	0.03	$&$	-0.98	\pm	0.26	$&$	-0.93	\pm	0.29	$&$	7.91	\pm	0.14	$&$	-1.66	\pm	0.21	$&$	-2.64	\pm	0.30	$\\
3C	300	&	HEG	&$	0.27	$&$	-0.08	\pm	0.14	$&$	-0.14	\pm	0.25	$&$	0.12	\pm	0.06	$&$	-0.63	\pm	0.19	$&$	-0.52	\pm	0.25	$&$	7.92	\pm	0.16	$&$	-1.55	\pm	0.22	$&$	-2.18	\pm	0.30	$\\
3C	438	&	HEG	&$	0.29	$&$	0.12	\pm	0.12	$&$	0.20	\pm	0.20	$&$	0.22	\pm	0.08	$&$	-0.40	\pm	0.12	$&$	-0.18	\pm	0.20	$&$	8.71	\pm	0.19	$&$	-2.14	\pm	0.24	$&$	-2.54	\pm	0.30	$\\
3C	299	&	HEG	&$	0.367	$&$	-0.18	\pm	0.14	$&$	-0.33	\pm	0.25	$&$	0.08	\pm	0.04	$&$	-0.79	\pm	0.21	$&$	-0.71	\pm	0.25	$&$	7.79	\pm	0.15	$&$	-1.37	\pm	0.21	$&$	-2.16	\pm	0.31	$\\
3C	42	&	HEG	&$	0.395	$&$	-0.05	\pm	0.13	$&$	-0.08	\pm	0.23	$&$	0.13	\pm	0.06	$&$	-0.59	\pm	0.17	$&$	-0.46	\pm	0.23	$&$	8.35	\pm	0.17	$&$	-1.90	\pm	0.22	$&$	-2.49	\pm	0.31	$\\
3C	16	&	HEG	&$	0.405	$&$	0.06	\pm	0.13	$&$	0.10	\pm	0.22	$&$	0.18	\pm	0.08	$&$	-0.46	\pm	0.14	$&$	-0.28	\pm	0.22	$&$	8.20	\pm	0.19	$&$	-1.73	\pm	0.24	$&$	-2.20	\pm	0.31	$\\
3C	274.1	&	HEG	&$	0.422	$&$	0.20	\pm	0.13	$&$	0.32	\pm	0.20	$&$	0.27	\pm	0.09	$&$	-0.33	\pm	0.10	$&$	-0.06	\pm	0.20	$&$	8.60	\pm	0.21	$&$	-2.06	\pm	0.25	$&$	-2.39	\pm	0.31	$\\
3C	457	&	HEG	&$	0.428	$&$	-0.37	\pm	0.13	$&$	-0.07	\pm	0.23	$&$	0.04	\pm	0.02	$&$	-1.09	\pm	0.23	$&$	-1.05	\pm	0.25	$&$	7.72	\pm	0.14	$&$	-1.20	\pm	0.20	$&$	-2.62	\pm	0.33	$\\
3C	244.1	&	HEG	&$	0.428	$&$	-0.04	\pm	0.13	$&$	-0.67	\pm	0.25	$&$	0.13	\pm	0.06	$&$	-0.58	\pm	0.17	$&$	-0.45	\pm	0.23	$&$	8.40	\pm	0.19	$&$	-2.04	\pm	0.23	$&$	-2.28	\pm	0.31	$\\
3C	46	&	HEG	&$	0.437	$&$	-0.24	\pm	0.13	$&$	-0.44	\pm	0.25	$&$	0.06	\pm	0.03	$&$	-0.88	\pm	0.21	$&$	-0.82	\pm	0.25	$&$	8.31	\pm	0.15	$&$	-1.82	\pm	0.21	$&$	-2.70	\pm	0.31	$\\
3C	341	&	HEG	&$	0.448	$&$	-0.19	\pm	0.13	$&$	-0.34	\pm	0.24	$&$	0.08	\pm	0.04	$&$	-0.80	\pm	0.20	$&$	-0.73	\pm	0.24	$&$	8.22	\pm	0.16	$&$	-1.72	\pm	0.22	$&$	-2.53	\pm	0.31	$\\
3C	172	&	HEG	&$	0.519	$&$	-0.18	\pm	0.13	$&$	-0.31	\pm	0.24	$&$	0.08	\pm	0.04	$&$	-0.77	\pm	0.20	$&$	-0.69	\pm	0.24	$&$	8.12	\pm	0.17	$&$	-1.58	\pm	0.22	$&$	-2.35	\pm	0.33	$\\
3C	330	&	HEG	&$	0.549	$&$	-0.14	\pm	0.13	$&$	-0.24	\pm	0.24	$&$	0.09	\pm	0.05	$&$	-0.72	\pm	0.19	$&$	-0.62	\pm	0.24	$&$	8.39	\pm	0.18	$&$	-1.65	\pm	0.24	$&$	-2.37	\pm	0.34	$\\
3C	49	&	HEG	&$	0.621	$&$	-0.11	\pm	0.13	$&$	0.05	\pm	0.20	$&$	0.10	\pm	0.05	$&$	-0.67	\pm	0.18	$&$	-0.57	\pm	0.23	$&$	8.44	\pm	0.18	$&$	-1.83	\pm	0.23	$&$	-2.50	\pm	0.33	$\\
3C	337	&	HEG	&$	0.635	$&$	0.01	\pm	0.13	$&$	-0.18	\pm	0.23	$&$	0.15	\pm	0.07	$&$	-0.53	\pm	0.16	$&$	-0.37	\pm	0.22	$&$	8.43	\pm	0.20	$&$	-1.99	\pm	0.24	$&$	-2.51	\pm	0.33	$\\
3C	34	&	HEG	&$	0.69	$&$	-0.32	\pm	0.12	$&$	0.01	\pm	0.22	$&$	0.05	\pm	0.02	$&$	-1.01	\pm	0.21	$&$	-0.96	\pm	0.23	$&$	8.21	\pm	0.16	$&$	-1.51	\pm	0.22	$&$	-2.51	\pm	0.34	$\\
3C	441	&	HEG	&$	0.708	$&$	-0.02	\pm	0.13	$&$	-0.58	\pm	0.23	$&$	0.14	\pm	0.06	$&$	-0.56	\pm	0.16	$&$	-0.42	\pm	0.22	$&$	8.68	\pm	0.20	$&$	-1.98	\pm	0.25	$&$	-2.54	\pm	0.34	$\\
3C	247	&	HEG	&$	0.749	$&$	-0.36	\pm	0.13	$&$	-0.04	\pm	0.22	$&$	0.04	\pm	0.02	$&$	-1.07	\pm	0.23	$&$	-1.03	\pm	0.26	$&$	8.34	\pm	0.16	$&$	-1.78	\pm	0.22	$&$	-2.85	\pm	0.34	$\\
3C	277.2	&	HEG	&$	0.766	$&$	-0.11	\pm	0.13	$&$	0.38	\pm	0.20	$&$	0.10	\pm	0.05	$&$	-0.67	\pm	0.18	$&$	-0.57	\pm	0.22	$&$	8.36	\pm	0.19	$&$	-1.60	\pm	0.24	$&$	-2.28	\pm	0.35	$\\
3C	340	&	HEG	&$	0.775	$&$	-0.03	\pm	0.13	$&$	-0.65	\pm	0.26	$&$	0.14	\pm	0.06	$&$	-0.56	\pm	0.16	$&$	-0.43	\pm	0.22	$&$	8.47	\pm	0.20	$&$	-1.77	\pm	0.25	$&$	-2.34	\pm	0.34	$\\
3C	352	&	HEG	&$	0.806	$&$	-0.17	\pm	0.13	$&$	-0.19	\pm	0.22	$&$	0.08	\pm	0.04	$&$	-0.77	\pm	0.19	$&$	-0.69	\pm	0.23	$&$	8.43	\pm	0.18	$&$	-1.67	\pm	0.23	$&$	-2.44	\pm	0.35	$\\
3C	263.1	&	HEG	&$	0.824	$&$	-0.02	\pm	0.13	$&$	-0.04	\pm	0.22	$&$	0.14	\pm	0.06	$&$	-0.56	\pm	0.16	$&$	-0.42	\pm	0.22	$&$	8.71	\pm	0.21	$&$	-1.86	\pm	0.26	$&$	-2.42	\pm	0.35	$\\
3C	217	&	HEG	&$	0.897	$&$	0.06	\pm	0.13	$&$	-0.30	\pm	0.23	$&$	0.18	\pm	0.07	$&$	-0.47	\pm	0.14	$&$	-0.29	\pm	0.21	$&$	8.35	\pm	0.23	$&$	-1.56	\pm	0.27	$&$	-2.03	\pm	0.36	$\\
3C	175.1	&	HEG	&$	0.92	$&$	0.08	\pm	0.13	$&$	-0.03	\pm	0.22	$&$	0.19	\pm	0.08	$&$	-0.45	\pm	0.14	$&$	-0.25	\pm	0.21	$&$	8.64	\pm	0.23	$&$	-1.82	\pm	0.28	$&$	-2.27	\pm	0.36	$\\
3C	289	&	HEG	&$	0.967	$&$	0.03	\pm	0.14	$&$	0.09	\pm	0.21	$&$	0.17	\pm	0.07	$&$	-0.50	\pm	0.15	$&$	-0.33	\pm	0.23	$&$	8.93	\pm	0.23	$&$	-2.17	\pm	0.28	$&$	-2.67	\pm	0.37	$\\
3C	280	&	HEG	&$	0.996	$&$	-0.44	\pm	0.14	$&$	0.13	\pm	0.21	$&$	0.03	\pm	0.02	$&$	-1.22	\pm	0.26	$&$	-1.19	\pm	0.27	$&$	8.21	\pm	0.17	$&$	-1.55	\pm	0.23	$&$	-2.77	\pm	0.37	$\\
3C	356	&	HEG	&$	1.079	$&$	0.08	\pm	0.15	$&$	0.05	\pm	0.23	$&$	0.20	\pm	0.09	$&$	-0.44	\pm	0.15	$&$	-0.24	\pm	0.24	$&$	9.01	\pm	0.25	$&$	-2.01	\pm	0.31	$&$	-2.45	\pm	0.39	$\\
3C	252	&	HEG	&$	1.103	$&$	-0.04	\pm	0.13	$&$	-0.81	\pm	0.27	$&$	0.13	\pm	0.06	$&$	-0.58	\pm	0.17	$&$	-0.45	\pm	0.23	$&$	8.72	\pm	0.22	$&$	-1.81	\pm	0.27	$&$	-2.39	\pm	0.38	$\\
3C	368	&	HEG	&$	1.132	$&$	-0.01	\pm	0.14	$&$	0.14	\pm	0.24	$&$	0.14	\pm	0.07	$&$	-0.55	\pm	0.17	$&$	-0.41	\pm	0.23	$&$	8.90	\pm	0.23	$&$	-1.91	\pm	0.28	$&$	-2.46	\pm	0.39	$\\
3C	267	&	HEG	&$	1.14	$&$	0.02	\pm	0.14	$&$	-0.06	\pm	0.23	$&$	0.16	\pm	0.07	$&$	-0.52	\pm	0.17	$&$	-0.36	\pm	0.24	$&$	8.87	\pm	0.24	$&$	-1.94	\pm	0.29	$&$	-2.45	\pm	0.39	$\\
3C	324	&	HEG	&$	1.206	$&$	-0.12	\pm	0.15	$&$	-0.02	\pm	0.23	$&$	0.10	\pm	0.06	$&$	-0.69	\pm	0.22	$&$	-0.59	\pm	0.27	$&$	8.87	\pm	0.22	$&$	-1.99	\pm	0.28	$&$	-2.68	\pm	0.40	$\\
3C	266	&	HEG	&$	1.275	$&$	-0.08	\pm	0.14	$&$	0.03	\pm	0.24	$&$	0.11	\pm	0.06	$&$	-0.64	\pm	0.19	$&$	-0.52	\pm	0.24	$&$	8.73	\pm	0.23	$&$	-1.77	\pm	0.28	$&$	-2.40	\pm	0.40	$\\
3C	13	&	HEG	&$	1.351	$&$	-0.05	\pm	0.14	$&$	-0.21	\pm	0.27	$&$	0.13	\pm	0.06	$&$	-0.59	\pm	0.18	$&$	-0.46	\pm	0.24	$&$	9.01	\pm	0.24	$&$	-2.01	\pm	0.29	$&$	-2.60	\pm	0.41	$\\
4C	13.66	&	HEG	&$	1.45	$&$	0.08	\pm	0.14	$&$	-0.14	\pm	0.24	$&$	0.19	\pm	0.08	$&$	-0.44	\pm	0.15	$&$	-0.25	\pm	0.23	$&$	8.75	\pm	0.27	$&$	-1.75	\pm	0.32	$&$	-2.20	\pm	0.42	$\\
3C	437	&	HEG	&$	1.48	$&$	0.31	\pm	0.15	$&$	-0.08	\pm	0.24	$&$	0.36	\pm	0.12	$&$	-0.25	\pm	0.09	$&$	0.10	\pm	0.21	$&$	9.13	\pm	0.32	$&$	-1.91	\pm	0.37	$&$	-2.16	\pm	0.43	$\\
3C	241	&	HEG	&$	1.617	$&$	0.03	\pm	0.15	$&$	0.13	\pm	0.23	$&$	0.16	\pm	0.08	$&$	-0.50	\pm	0.17	$&$	-0.34	\pm	0.25	$&$	9.07	\pm	0.27	$&$	-1.99	\pm	0.32	$&$	-2.49	\pm	0.44	$\\
3C	470	&	HEG	&$	1.653	$&$	0.19	\pm	0.15	$&$	0.49	\pm	0.21	$&$	0.27	\pm	0.11	$&$	-0.34	\pm	0.12	$&$	-0.07	\pm	0.23	$&$	9.11	\pm	0.31	$&$	-2.08	\pm	0.35	$&$	-2.42	\pm	0.44	$\\
3C	322	&	HEG	&$	1.681	$&$	0.29	\pm	0.16	$&$	-0.08	\pm	0.27	$&$	0.34	\pm	0.12	$&$	-0.27	\pm	0.10	$&$	0.07	\pm	0.23	$&$	9.24	\pm	0.33	$&$	-2.09	\pm	0.38	$&$	-2.36	\pm	0.45	$\\
3C	239	&	HEG	&$	1.781	$&$	0.17	\pm	0.16	$&$	0.04	\pm	0.25	$&$	0.25	\pm	0.11	$&$	-0.36	\pm	0.14	$&$	-0.11	\pm	0.26	$&$	9.21	\pm	0.32	$&$	-2.07	\pm	0.37	$&$	-2.43	\pm	0.46	$\\
3C	294	&	HEG	&$	1.786	$&$	0.05	\pm	0.16	$&$	0.31	\pm	0.23	$&$	0.18	\pm	0.09	$&$	-0.48	\pm	0.17	$&$	-0.30	\pm	0.26	$&$	8.98	\pm	0.29	$&$	-1.87	\pm	0.34	$&$	-2.34	\pm	0.46	$\\
3C	225B	&	HEG	&$	0.582	$&$	0.03	\pm	0.12	$&$	0.45	\pm	0.23	$&$	0.17	\pm	0.07	$&$	-0.50	\pm	0.14	$&$	-0.33	\pm	0.21	$&$	8.50	\pm	0.20	$&$	-1.77	\pm	0.25	$&$	-2.27	\pm	0.33	$\\
3C	55	&	HEG	&$	0.735	$&$	0.24	\pm	0.13	$&$	0.27	\pm	0.26	$&$	0.30	\pm	0.10	$&$	-0.30	\pm	0.10	$&$	0.00	\pm	0.20	$&$	8.86	\pm	0.24	$&$	-1.94	\pm	0.29	$&$	-2.24	\pm	0.35	$\\
3C	68.2	&	HEG	&$	1.575	$&$	-0.04	\pm	0.16	$&$	0.08	\pm	0.26	$&$	0.13	\pm	0.07	$&$	-0.59	\pm	0.20	$&$	-0.46	\pm	0.27	$&$	8.95	\pm	0.25	$&$	-2.00	\pm	0.30	$&$	-2.59	\pm	0.44	$\\

\hline																																		
\hline																																		
\end{tabular}																																		
\end{minipage}																																		
\end{table*}																																		
																																		
 																																		
\begin{table*}																																		
\begin{minipage}{165mm}																																		
\scriptsize																																		
\caption{Outflow and Spin Properties for FRII LEG, Q, \& W Sources}																																		
\label{FRIIsources-LEG-plus W sources new-paper}																																		
\begin{tabular}{lllllllllll}   
\hline\hline  																																		
(1)&(2)&(3)&(4)&(5)&(6)&(7)&(8)&(9)&(10)&(11)\\																																		
Source&Type&z&$\rm{Log}$ &	$\rm{Log}$&																																	
$\rm{Log}$&$\rm{Log}$&$\rm{Log}$&$\rm{Log(\rm{M_{spin}})}$&$\rm{Log}$&$\rm{Log}$\\																																		
&&&$\rm{(F)}$&$\rm{(\rm{E_{spin}}/E_{spin.max})}$&
$\rm{(\rm{M_{dyn}}/\rm{M_{irr}})}$&$\rm{(\rm{M_{spin}}/\rm{M_{dyn}})}$&$\rm{(\rm{M_{spin}}/\rm{M_{irr}})}$&${(M_{\odot})}$&$\rm{(E_T/\rm{E_{spin}})} $&$\rm{(E_T/M_{dyn})}$\\																																		
\hline																							
3C	35	&	LEG	&$	0.067	$&$	-0.25	\pm	0.22	$&$	-0.46	\pm	0.41	$&$	0.06	\pm	0.05	$&$	-0.90	\pm	0.36	$&$	-0.84	\pm	0.41	$&$	7.75	\pm	0.14	$&$	-2.01	\pm	0.26	$&$	-2.91	\pm	0.33	$\\
3C	326	&	LEG	&$	0.088	$&$	-0.09	\pm	0.19	$&$	-0.16	\pm	0.33	$&$	0.11	\pm	0.07	$&$	-0.65	\pm	0.26	$&$	-0.54	\pm	0.33	$&$	7.73	\pm	0.15	$&$	-1.77	\pm	0.24	$&$	-2.42	\pm	0.31	$\\
3C	236	&	LEG	&$	0.099	$&$	-0.20	\pm	0.19	$&$	-0.36	\pm	0.35	$&$	0.07	\pm	0.05	$&$	-0.82	\pm	0.30	$&$	-0.75	\pm	0.35	$&$	7.99	\pm	0.14	$&$	-2.06	\pm	0.24	$&$	-2.88	\pm	0.32	$\\
4C	12.03	&	LEG	&$	0.156	$&$	-0.19	\pm	0.18	$&$	-0.33	\pm	0.32	$&$	0.08	\pm	0.05	$&$	-0.79	\pm	0.27	$&$	-0.71	\pm	0.32	$&$	8.06	\pm	0.14	$&$	-2.01	\pm	0.22	$&$	-2.80	\pm	0.31	$\\
3C	319	&	LEG	&$	0.192	$&$	0.10	\pm	0.16	$&$	0.17	\pm	0.25	$&$	0.21	\pm	0.10	$&$	-0.42	\pm	0.16	$&$	-0.21	\pm	0.25	$&$	7.90	\pm	0.18	$&$	-1.69	\pm	0.24	$&$	-2.11	\pm	0.30	$\\
3C	132	&	LEG	&$	0.214	$&$	-0.17	\pm	0.15	$&$	-0.31	\pm	0.28	$&$	0.08	\pm	0.05	$&$	-0.77	\pm	0.23	$&$	-0.69	\pm	0.28	$&$	7.97	\pm	0.15	$&$	-1.74	\pm	0.22	$&$	-2.51	\pm	0.30	$\\
3C	123	&	LEG	&$	0.218	$&$	0.31	\pm	0.12	$&$	0.49	\pm	0.17	$&$	0.36	\pm	0.09	$&$	-0.25	\pm	0.07	$&$	0.11	\pm	0.17	$&$	8.57	\pm	0.21	$&$	-1.86	\pm	0.26	$&$	-2.11	\pm	0.30	$\\
3C	153	&	LEG	&$	0.277	$&$	-0.21	\pm	0.14	$&$	-0.38	\pm	0.26	$&$	0.07	\pm	0.04	$&$	-0.83	\pm	0.22	$&$	-0.76	\pm	0.26	$&$	8.11	\pm	0.15	$&$	-1.76	\pm	0.21	$&$	-2.59	\pm	0.30	$\\
4C	14.27	&	LEG	&$	0.392	$&$	0.03	\pm	0.14	$&$	0.06	\pm	0.23	$&$	0.17	\pm	0.07	$&$	-0.49	\pm	0.15	$&$	-0.33	\pm	0.23	$&$	8.24	\pm	0.19	$&$	-1.79	\pm	0.24	$&$	-2.28	\pm	0.31	$\\
3C	200	&	LEG	&$	0.458	$&$	-0.09	\pm	0.13	$&$	-0.15	\pm	0.23	$&$	0.11	\pm	0.05	$&$	-0.65	\pm	0.18	$&$	-0.54	\pm	0.23	$&$	8.30	\pm	0.17	$&$	-1.79	\pm	0.22	$&$	-2.43	\pm	0.31	$\\
3C	295	&	LEG	&$	0.461	$&$	0.23	\pm	0.12	$&$	0.37	\pm	0.18	$&$	0.29	\pm	0.09	$&$	-0.31	\pm	0.09	$&$	-0.02	\pm	0.18	$&$	9.15	\pm	0.22	$&$	-2.29	\pm	0.27	$&$	-2.60	\pm	0.32	$\\
3C	19	&	LEG	&$	0.482	$&$	-0.14	\pm	0.13	$&$	-0.24	\pm	0.23	$&$	0.09	\pm	0.05	$&$	-0.72	\pm	0.19	$&$	-0.63	\pm	0.23	$&$	8.44	\pm	0.17	$&$	-1.91	\pm	0.22	$&$	-2.63	\pm	0.32	$\\
3C	427.1	&	LEG	&$	0.572	$&$	-0.24	\pm	0.13	$&$	-0.44	\pm	0.24	$&$	0.06	\pm	0.03	$&$	-0.88	\pm	0.21	$&$	-0.82	\pm	0.24	$&$	8.28	\pm	0.16	$&$	-1.74	\pm	0.21	$&$	-2.62	\pm	0.33	$\\
3C	249.1	&	Q	&$	0.311	$&$	-0.56	\pm	0.17	$&$	-1.04	\pm	0.33	$&$	0.02	\pm	0.01	$&$	-1.44	\pm	0.31	$&$	-1.42	\pm	0.33	$&$	7.86	\pm	0.19	$&$	-1.53	\pm	0.24	$&$	-2.96	\pm	0.43	$\\
3C	351	&	Q	&$	0.371	$&$	-0.38	\pm	0.16	$&$	-0.70	\pm	0.30	$&$	0.04	\pm	0.02	$&$	-1.12	\pm	0.28	$&$	-1.08	\pm	0.30	$&$	8.38	\pm	0.20	$&$	-1.94	\pm	0.25	$&$	-3.05	\pm	0.43	$\\
3C	215	&	Q	&$	0.411	$&$	0.04	\pm	0.16	$&$	0.07	\pm	0.26	$&$	0.17	\pm	0.09	$&$	-0.49	\pm	0.18	$&$	-0.32	\pm	0.26	$&$	7.81	\pm	0.27	$&$	-1.33	\pm	0.31	$&$	-1.82	\pm	0.43	$\\
3C	47	&	Q	&$	0.425	$&$	-0.20	\pm	0.15	$&$	-0.36	\pm	0.27	$&$	0.07	\pm	0.04	$&$	-0.82	\pm	0.23	$&$	-0.75	\pm	0.27	$&$	8.38	\pm	0.22	$&$	-1.74	\pm	0.26	$&$	-2.56	\pm	0.42	$\\
3C	334	&	Q	&$	0.555	$&$	-0.31	\pm	0.15	$&$	-0.57	\pm	0.28	$&$	0.05	\pm	0.03	$&$	-0.99	\pm	0.25	$&$	-0.95	\pm	0.28	$&$	8.71	\pm	0.21	$&$	-2.01	\pm	0.26	$&$	-3.01	\pm	0.43	$\\
3C	275.1	&	Q	&$	0.557	$&$	-0.26	\pm	0.15	$&$	-0.46	\pm	0.29	$&$	0.06	\pm	0.04	$&$	-0.90	\pm	0.25	$&$	-0.84	\pm	0.29	$&$	7.40	\pm	0.22	$&$	-1.13	\pm	0.26	$&$	-2.03	\pm	0.42	$\\
3C	263	&	Q	&$	0.646	$&$	-0.20	\pm	0.14	$&$	-0.35	\pm	0.27	$&$	0.07	\pm	0.04	$&$	-0.81	\pm	0.23	$&$	-0.74	\pm	0.27	$&$	8.29	\pm	0.23	$&$	-1.58	\pm	0.27	$&$	-2.39	\pm	0.43	$\\
3C	207	&	Q	&$	0.684	$&$	0.12	\pm	0.14	$&$	0.20	\pm	0.23	$&$	0.22	\pm	0.09	$&$	-0.40	\pm	0.14	$&$	-0.18	\pm	0.23	$&$	8.10	\pm	0.29	$&$	-1.38	\pm	0.33	$&$	-1.78	\pm	0.43	$\\
3C	254	&	Q	&$	0.734	$&$	-0.27	\pm	0.15	$&$	-0.49	\pm	0.29	$&$	0.06	\pm	0.03	$&$	-0.93	\pm	0.26	$&$	-0.87	\pm	0.29	$&$	8.37	\pm	0.21	$&$	-1.68	\pm	0.26	$&$	-2.60	\pm	0.43	$\\
3C	175	&	Q	&$	0.768	$&$	-0.15	\pm	0.15	$&$	-0.27	\pm	0.26	$&$	0.09	\pm	0.05	$&$	-0.74	\pm	0.22	$&$	-0.65	\pm	0.27	$&$	9.16	\pm	0.23	$&$	-2.31	\pm	0.28	$&$	-3.05	\pm	0.43	$\\
3C	196	&	Q	&$	0.871	$&$	0.19	\pm	0.15	$&$	0.31	\pm	0.23	$&$	0.27	\pm	0.11	$&$	-0.34	\pm	0.12	$&$	-0.07	\pm	0.23	$&$	9.26	\pm	0.31	$&$	-2.16	\pm	0.36	$&$	-2.50	\pm	0.44	$\\
3C	309.1	&	Q	&$	0.904	$&$	-0.02	\pm	0.14	$&$	-0.03	\pm	0.25	$&$	0.14	\pm	0.07	$&$	-0.56	\pm	0.18	$&$	-0.41	\pm	0.25	$&$	8.54	\pm	0.26	$&$	-1.66	\pm	0.30	$&$	-2.21	\pm	0.43	$\\
3C	336	&	Q	&$	0.927	$&$	-0.08	\pm	0.14	$&$	-0.14	\pm	0.25	$&$	0.11	\pm	0.06	$&$	-0.64	\pm	0.20	$&$	-0.53	\pm	0.25	$&$	8.56	\pm	0.25	$&$	-1.76	\pm	0.29	$&$	-2.40	\pm	0.43	$\\
3C	245	&	Q	&$	1.029	$&$	-0.06	\pm	0.14	$&$	-0.10	\pm	0.25	$&$	0.12	\pm	0.06	$&$	-0.61	\pm	0.19	$&$	-0.48	\pm	0.25	$&$	8.79	\pm	0.25	$&$	-1.90	\pm	0.30	$&$	-2.51	\pm	0.43	$\\
3C	212	&	Q	&$	1.049	$&$	0.04	\pm	0.14	$&$	0.07	\pm	0.24	$&$	0.17	\pm	0.08	$&$	-0.49	\pm	0.16	$&$	-0.32	\pm	0.24	$&$	8.71	\pm	0.27	$&$	-1.79	\pm	0.31	$&$	-2.27	\pm	0.43	$\\
3C	186	&	Q	&$	1.063	$&$	-0.10	\pm	0.14	$&$	-0.17	\pm	0.26	$&$	0.11	\pm	0.06	$&$	-0.66	\pm	0.20	$&$	-0.56	\pm	0.26	$&$	8.84	\pm	0.24	$&$	-1.89	\pm	0.29	$&$	-2.55	\pm	0.43	$\\
3C	208	&	Q	&$	1.11	$&$	0.04	\pm	0.15	$&$	0.06	\pm	0.24	$&$	0.17	\pm	0.08	$&$	-0.49	\pm	0.16	$&$	-0.32	\pm	0.24	$&$	8.91	\pm	0.27	$&$	-1.93	\pm	0.32	$&$	-2.42	\pm	0.43	$\\
3C	204	&	Q	&$	1.112	$&$	0.00	\pm	0.14	$&$	0.00	\pm	0.24	$&$	0.15	\pm	0.07	$&$	-0.54	\pm	0.17	$&$	-0.39	\pm	0.24	$&$	8.96	\pm	0.26	$&$	-2.06	\pm	0.30	$&$	-2.59	\pm	0.43	$\\
3C	190	&	Q	&$	1.197	$&$	0.20	\pm	0.15	$&$	0.32	\pm	0.22	$&$	0.27	\pm	0.10	$&$	-0.33	\pm	0.12	$&$	-0.06	\pm	0.22	$&$	8.37	\pm	0.31	$&$	-1.38	\pm	0.35	$&$	-1.71	\pm	0.43	$\\
3C	68.1	&	Q	&$	1.238	$&$	0.06	\pm	0.15	$&$	0.09	\pm	0.25	$&$	0.18	\pm	0.08	$&$	-0.47	\pm	0.16	$&$	-0.29	\pm	0.25	$&$	9.43	\pm	0.28	$&$	-2.33	\pm	0.33	$&$	-2.80	\pm	0.44	$\\
4C	16.49	&	Q	&$	1.296	$&$	-0.06	\pm	0.14	$&$	-0.10	\pm	0.25	$&$	0.13	\pm	0.06	$&$	-0.60	\pm	0.19	$&$	-0.48	\pm	0.25	$&$	9.20	\pm	0.25	$&$	-2.23	\pm	0.30	$&$	-2.83	\pm	0.43	$\\
3C	181	&	Q	&$	1.382	$&$	-0.14	\pm	0.15	$&$	-0.25	\pm	0.27	$&$	0.09	\pm	0.05	$&$	-0.72	\pm	0.22	$&$	-0.63	\pm	0.27	$&$	8.88	\pm	0.24	$&$	-1.82	\pm	0.30	$&$	-2.55	\pm	0.44	$\\
3C	268.4	&	Q	&$	1.4	$&$	0.03	\pm	0.18	$&$	0.05	\pm	0.30	$&$	0.17	\pm	0.10	$&$	-0.50	\pm	0.20	$&$	-0.33	\pm	0.30	$&$	9.30	\pm	0.27	$&$	-2.05	\pm	0.34	$&$	-2.55	\pm	0.45	$\\
3C	14	&	Q	&$	1.469	$&$	-0.04	\pm	0.15	$&$	-0.07	\pm	0.25	$&$	0.13	\pm	0.07	$&$	-0.59	\pm	0.19	$&$	-0.46	\pm	0.25	$&$	8.81	\pm	0.26	$&$	-1.82	\pm	0.31	$&$	-2.41	\pm	0.43	$\\
3C	270.1	&	Q	&$	1.519	$&$	0.41	\pm	0.15	$&$	0.62	\pm	0.21	$&$	0.44	\pm	0.13	$&$	-0.20	\pm	0.08	$&$	0.24	\pm	0.21	$&$	8.80	\pm	0.34	$&$	-1.56	\pm	0.39	$&$	-1.76	\pm	0.44	$\\
3C	205	&	Q	&$	1.534	$&$	-0.06	\pm	0.15	$&$	-0.10	\pm	0.26	$&$	0.12	\pm	0.06	$&$	-0.61	\pm	0.19	$&$	-0.49	\pm	0.26	$&$	8.99	\pm	0.25	$&$	-1.94	\pm	0.30	$&$	-2.55	\pm	0.44	$\\
3C	432	&	Q	&$	1.805	$&$	-0.07	\pm	0.15	$&$	-0.11	\pm	0.26	$&$	0.12	\pm	0.06	$&$	-0.62	\pm	0.20	$&$	-0.50	\pm	0.26	$&$	9.48	\pm	0.25	$&$	-2.37	\pm	0.31	$&$	-2.98	\pm	0.44	$\\
3C	191	&	Q	&$	1.956	$&$	-0.01	\pm	0.15	$&$	-0.01	\pm	0.26	$&$	0.15	\pm	0.08	$&$	-0.54	\pm	0.19	$&$	-0.39	\pm	0.26	$&$	9.16	\pm	0.26	$&$	-1.98	\pm	0.32	$&$	-2.52	\pm	0.44	$\\
3C	9	&	Q	&$	2.012	$&$	0.14	\pm	0.16	$&$	0.24	\pm	0.25	$&$	0.23	\pm	0.11	$&$	-0.38	\pm	0.15	$&$	-0.15	\pm	0.25	$&$	9.42	\pm	0.30	$&$	-2.14	\pm	0.36	$&$	-2.52	\pm	0.45	$\\
3C	223	&	W	&$	0.136	$&$	-0.36	\pm	0.17	$&$	-0.65	\pm	0.33	$&$	0.04	\pm	0.03	$&$	-1.07	\pm	0.30	$&$	-1.03	\pm	0.33	$&$	7.41	\pm	0.13	$&$	-1.34	\pm	0.22	$&$	-2.41	\pm	0.31	$\\
3C	79	&	W	&$	0.255	$&$	-0.12	\pm	0.13	$&$	-0.22	\pm	0.24	$&$	0.10	\pm	0.05	$&$	-0.70	\pm	0.19	$&$	-0.60	\pm	0.24	$&$	8.09	\pm	0.16	$&$	-1.63	\pm	0.22	$&$	-2.33	\pm	0.30	$\\
3C	109	&	W	&$	0.305	$&$	-0.14	\pm	0.16	$&$	-0.25	\pm	0.28	$&$	0.09	\pm	0.05	$&$	-0.72	\pm	0.23	$&$	-0.63	\pm	0.28	$&$	7.58	\pm	0.24	$&$	-1.12	\pm	0.28	$&$	-1.84	\pm	0.43	$\\
\hline																																		
\hline																																		
\end{tabular}																																		
\end{minipage}																																		
\end{table*}

\begin{table*}
\begin{minipage}{165mm}
\scriptsize
\caption{Unweighted Mean Value and Standard Deviation of Black Hole Spin Mass-Energy Parameters (top five rows; values in parentheses indicate the average uncertainty per source), and Unweighted Best Fit Values to Each Quantity vs. Log(1+z) (bottom three rows).}   
\label{tab:mean}        
\begin{tabular}{lllllllll}   
\hline\hline                    
(1)&(2)&(3)&(4)&(5)&(6)&(7)&(8)&(9)\\
	&		&$\rm{Log}$	&$\rm{Log}$&$\rm{Log}$&$\rm{Log}$&$	\rm{Log}$&$	\rm{Log}$&$	\rm{Log}$\\
	
		&N			&$\rm{(\rm{M_{dyn}}/\rm{M_{irr}})}$&$	\rm{({M_{spin}}/{M_{dyn}})}$&$(\rm{M_{spin}/M_{irr}})$\footnote{$\rm{Log(E_{spin}/E_{spin,max}) \simeq 0.38 + Log(M_{spin}/M_{irr})}$ since 
		$\rm{(E_{spin}/E_{spin,max}) \simeq 2.41 
	(M_{spin}/M_{irr})}$. This does not affect the standard deviation or average uncertainty per source listed in the top five rows of the table; thus the unweighted mean value of $\rm{Log(E_{spin}/E_{spin,max})}$ is obtained by adding 0.38 to 
	that listed for $\rm{Log(M_{spin}/M_{irr})}$, bringing the value for 100 sources to $-0.15  \pm	0.33(0.26)$, for example. 
	It does not affect the slope or $\chi^2$ listed in the bottom part of the table for $\rm{Log(M_{spin}/M_{irr}})$ though it does add 0.38 to the y-intercept, bringing this value to -0.42 for $\rm{Log(E_{spin}/E_{spin,max})}$. }	&$(\rm{\rm{M_{spin}}/M_{\odot})}$&(F)&$\rm{(E_{T}/\rm{E_{spin}})}$&$\rm{(E_{T}/\rm{M_{dyn}})}$	\\
\hline
HEG	&55	&$	0.13 \pm	0.08 (0.06)	$&$	-0.66	\pm	0.27 (0.19)	$&$	-0.53	\pm	0.34(0.25)	$&$	8.41	\pm	0.50(0.25)	$&$-0.08 \pm 0.19 (0.14) $&$	-1.80	\pm	0.23 (0.30)	$&$-2.46 \pm 0.19 (0.35) $	\\
Q	&29	&$	0.14 \pm	0.09 (0.07)	$&$	-0.64 	\pm	0.26 (0.19)	$&$	-0.50	\pm	0.34(0.26)	$&$	8.72	\pm	0.52(0.26)	$&$-0.06 \pm 0.19 (0.15) $&$	-1.84	\pm	0.31 (0.31)	$&$-2.48 \pm 0.38(0.43) $	\\
LEG	&13	&$	0.14	\pm	0.10 (0.06)	$&$	-0.65	\pm	0.22 (0.21)	$&$	-0.52	\pm	0.31(0.27)	$&$	8.19	\pm	0.38(0.27)	$&$-0.07 \pm 0.18 (0.15) $&$	-1.88	\pm	0.17 (0.31)	$&$-2.53 \pm 0.26(0.31) $	\\
W	&3	&$	0.08 \pm	0.03 (0.04)	$&$	-0.83 \pm	0.21 (0.24)	$&$	-0.75	\pm	0.24(0.28)	$&$	7.69	\pm	0.35 (0.28)	$&$-0.21 \pm 0.13(0.15)$&$	-1.37	\pm	0.26 (0.32)	$&$-2.19 \pm 0.31(0.34) $	\\
All	&100	&$	0.13	\pm	0.08 (0.06)	$&$	-0.66	\pm	0.26(0.20)	$&$	-0.53	\pm	0.33(0.26)	$&$	8.45	\pm	0.53 (0.26)	$&$-0.08 \pm 0.19 (0.15) $&$	-1.81	\pm	0.26 (0.30)	$&$-2.47 \pm 0.27(0.37) $	\\
\hline 
Slope&100& $0.28 \pm 0.06$&$0.97 \pm 0.18 $&$1.25 \pm 0.23 $&$3.46 \pm 0.23 $& $0.71 \pm 0.13 $&$	-0.73	\pm	0.19$&$0.25	\pm	0.22$\\
Y-int&100& $0.07 \pm 0.02 $&$-0.88 \pm 0.05 $&
$-0.80 \pm 0.06 $&$7.68 \pm 0.06 $&$ -0.23 \pm 0.03$&$	-1.65	\pm	0.05	$&$	-2.52	\pm	0.06$\\
$\chi^2 $&100& 0.54&$4.96 $&$8.33 $&$8.52 $&$2.71$&$5.95$&$7.27$ \\
\hline
\hline 
\end{tabular}
\end{minipage}
\end{table*}
																
 
\bsp	
\label{lastpage}
\end{document}